\documentclass{pasj01}
\draft
\Received{$\langle$reception date$\rangle$}
\Accepted{$\langle$acception date$\rangle$}
\Published{$\langle$publication date$\rangle$}

\newcommand{\HII}{H{\sc ii}}

\newcommand{\Msun}{{\it M}_{\odot}}

\usepackage{setspace}
\begin{document}

\title{Triggered high-mass star formation in the \HII\ region W28A2: a cloud-cloud collision scenario}

\author{Katsuhiro Hayashi$^{1,11}$, Satoshi Yoshiike$^{1}$, Rei Enokiya$^{1}$, Shinji Fujita$^{1,4}$, Rin Yamada$^{1}$, Hidetoshi Sano$^{1,2,12}$, Kazufumi Torii$^{3}$, Mikito Kohno$^{1,13}$, Atsushi Nishimura$^{4}$, Akio Ohama$^{1}$, Hiroaki Yamamoto$^{1}$, Kengo Tachihara$^{1}$, Graeme Wong$^{5,8}$, Nigel Maxted$^{6,7}$, Catherine Braiding$^{8}$, Gavin Rowell$^{9}$, Michael Burton$^{10}$ and Yasuo Fukui$^{1,2}$}%
\altaffiltext{1}{Department of Physics, Nagoya University, Chikusa-ku, Nagoya, Aichi, 464-8602, Japan}
\altaffiltext{2}{Institute for Advanced Research, Nagoya University, Chikusa-ku, Nagoya, Aichi, 464-8601, Japan}
\altaffiltext{3}{Nobeyama Radio Observatory, 462-2 Nobeyama Minamimaki-mura, Minamisaku-gun, Nagano 384-1305, Japan}
\altaffiltext{4}{Department of Physical Science, Graduate School of Science, Osaka Prefecture University, 1-1 Gakuen-cho, Naka-ku, Sakai, Osaka 599-8531, Japan}
\altaffiltext{5}{Pawsey Supercomputing Centre, 26 Dick Perry Ave, Kensington 6151, WA, Australia}
\altaffiltext{6}{Western Sydney University, Locked Bag 1797, Penrith, NSW 2751, Australia}
\altaffiltext{7}{School of Science, The University of New South Wales, Australian Defence Force Academy, Canberra 2610, Australia}
\altaffiltext{8}{School of Physics, The University of New South Wales, Sydney 2052, Australia}
\altaffiltext{9}{School of Physical Sciences, The University of Adelaide, Adelaide 5005, Australia}
\altaffiltext{10}{Armagh Observatory and Planetarium, College Hill, Armagh BT61 9DG, UK}
\altaffiltext{11}{Institute of Space and Astronautical Science (ISAS) Japan Aerospace Exploration Agency (JAXA), 3-1-1 Yoshinodai, Chuo-ku, Sagamihara, Kanagawa 252-5210, Japan}
\altaffiltext{12}{National Astronomical Observatory of Japan, Mitaka, Tokyo 181-8588, Japan}
\email{khayashi@a.phys.nagoya-u.ac.jp}
\altaffiltext{13}{Astronomy Section, Nagoya City Science Museum, 2-17-1 Sakae, Naka-ku, Nagoya, Aichi 460-0008, Japan}


\maketitle

\begin{abstract}
We report on a study of the high-mass star formation in the the \HII\ region W28A2 by investigating the molecular clouds extended over $\sim$5--10 pc from the exciting stars using the $^{12}$CO and $^{13}$CO~($J$$=$1--0) and $^{12}$CO~($J$$=$2--1) data taken by the NANTEN2 and Mopra observations. 
These molecular clouds consist of three velocity components with the CO intensity peaks at $V_{\rm LSR} \sim$ $-4$~km~s$^{-1}$, 9~km~s$^{-1}$ and 16~km~s$^{-1}$.
The highest CO intensity is detected at $V_{\rm LSR} \sim$$9$~km~s$^{-1}$, where the high-mass stars with the spectral types of O6.5--B0.5 are embedded. 
We found bridging features connecting these clouds toward the directions of the exciting sources.
Comparisons of the gas distributions with the radio continuum emission and 8 $\mu$m infrared emission show spatial coincidence/anti-coincidence, suggesting physical associations between the gas and the exciting sources.
The $^{12}$CO $J$$=$2--1 to 1--0 intensity ratio shows a high value ($\gtrsim$ 0.8) toward the exciting sources for the $-4$~km~s$^{-1}$ and $+9$~km~s$^{-1}$ clouds, possibly due to heating by the high-mass stars, whereas the intensity ratio at the CO intensity peak ($V_{\rm LSR} \sim$9~km~s$^{-1}$) lowers down to $\sim$0.6, suggesting self absorption by the dense gas in the near side of the $+9$~km~s$^{-1}$ cloud.
We found partly complementary gas distributions between the $-4$~km~s$^{-1}$ and $+9$~km~s$^{-1}$ clouds, and the $-4$~km~s$^{-1}$ and $+16$~km~s$^{-1}$ clouds.
The exciting sources are located toward the overlapping region in the $-4$~km~s$^{-1}$ and $+9$~km~s$^{-1}$ clouds.
Similar gas properties are found in the Galactic massive star clusters, RCW~38 and NGC~6334, where an early stage of cloud collision to trigger the star formation is suggested.
Based on these results, we discuss a possibility of the formation of high-mass stars in the W28A2 region triggered by the cloud-cloud collision.

\end{abstract}


\section{Introduction}
\label{sec:introduction}

\subsection{High-mass star formation}
\label{sec: Triggered high-mass star formation}

High-mass stars greatly influence physical and chemical environments of the interstellar medium (ISM) by injecting large energy through stellar winds, ultraviolet radiation (UV) and supernovae explosions at the end of their lives.
Materials around the high-mass stars are ionized by the UV photons, which make various sizes and shapes of the \HII\ regions.
Heavy elements produced by the supernova explosions affect the chemical evolution of the ISM. 
Revealing the formation process of high-mass stars is essential to understand physical properties of the ISM and evolution of galaxies. 

The early stage of star formation is generally explained by a contraction of the interstellar gas due to the self-gravity in a turbulent medium. 
Theoretically, ``core accretion'' and ``competitive accretion'' are commonly invoked scenarios to explain the formation of massive stars (e.g., for reviews, \cite{ZinneckerYorke07}; \cite{Tan+14}). 
While these models assume a gravitationally bound system, possible scenarios of external agents to trigger the formation of high-mass stars have been discussed (e.g., \cite{Elmegreen98}).
One of the triggering scenarios is expanding motion of the ionized gas (``collect \& collapse"; e.g., \cite{ElmegreenLada77}). 
The external pressure in a shock wave accumulates the surrounding material and forms gravitationally unstable dense cores, which collapses to create the next-generation of stars.
The other is collisions of molecular clouds (``cloud-cloud collision''; e.g., \cite{Loren79}; \cite{HabeOhta92}).
Incidental collisions of two clouds at a supersonic relative velocity accumulate the gas in a shock wave and generate gravitationally unstable dense cores, leading the formation of high-mass stars. 

A number of discoveries of Spitzer bubbles in the 2000 era (\cite{Churchwell+06}; \yearcite{Churchwell+07}) have been established the collect \& collapse as a plausible model to trigger the high-mass star formation (e.g., Sh~104: \cite{Deharveng+03}; RCW~79: \cite{Zavagno+05}). 
The size of the \HII\ regions in their bubbles can be explained by the UV radiation from the central massive stars, which promote the next generation of stars at the peripheries of the \HII\ region.
This model is also supported in terms of a theoretical aspect to form the bubble structure \citep{HosokawaInutsuka06}.
However, a clear diagnostics of the expanding gas motion by the ionized gas has not been found.
The pressure from the \HII\ region is easily to escape from the surface of the molecular gas because the shape of the molecular clouds is flattened rather than symmetric \citep{BeaumontWilliams10}.
Even if collect \& collapse can be applied to the sequential star formation, it does not explain the formation of the first born central massive stars. 

On the other hand, the model of cloud-cloud collision is easier to explain these problems.
The collision induces formations of dense clumps at the compressed region, achieving the large mass accretion rate (10$^{-4}$ to 10$^{-3}$ ${\it M}_{\odot}$~yr$^{-1}$) enough to create massive stars (e.g., \cite{InoueFukui13}; \cite{Takahira+14}). 
If the collision occurs between clouds with different sizes along the line of sight, the smaller cloud creates a hole in the larger cloud, which forms a ring-like gas distribution (e.g., Figure 12 in \cite{Torii+17}).
Depending on the angle and the elapsed time of the collision, the smaller cloud is displaced relative to the hole in the larger cloud.
The UV radiation from the massive stars ionizes the surrounding gas and creates the infrared ring associated with the gas distribution (e.g., \cite{Torii+15}).  
Unless the clouds are not dispersal by the ionization, these clouds show a complementary gas distribution.
The momentum exchange between the colliding clouds generates bridging structures in the position-velocity diagram.
Such cloud properties are found in several Galactic massive star clusters (e.g., Westerlund~2: \cite{Furukawa+09}; \cite{Ohama+10}; RCW~38: \cite{Fukui+16}; M42: \cite{Fukui+18a}; NGC~6334/NGC~6357: \cite{Fukui+18b}; NGC~6618: \cite{Nishimura+18}) and \HII\ regions which harbors a single or a few high-mass star(s) (e.g., RCW~120: \cite{Torii+15}; G35.20-0.74: \cite{Dewangan17}; RCW~32: \cite{Enokiya+18}; RCW~36: \cite{Sano+18}; RCW~34: \cite{Hayashi+18}; S44: \cite{Kohno+18}; N4: \cite{Fujita+19}), as well as the active star-forming regions in the Large Magellanic Cloud and M33 (e.g., \cite{Tachihara+18}; \cite{Tsuge+19}; \cite{Sano+19}).
\citet{Enokiya+19} performed a statistical study using these observational findings and found that the peak gas column density becomes larger as increasing the relative velocity between the colliding clouds in the Galactic disk.

\subsection{The \HII\ complex W28A2}
\label{sec:W28A2}

\begin{figure}[h]
 \begin{tabular}{cc}
  \begin{minipage}{0.5\hsize}
   \begin{center}
    \rotatebox{0}{\resizebox{9cm}{!}{\includegraphics{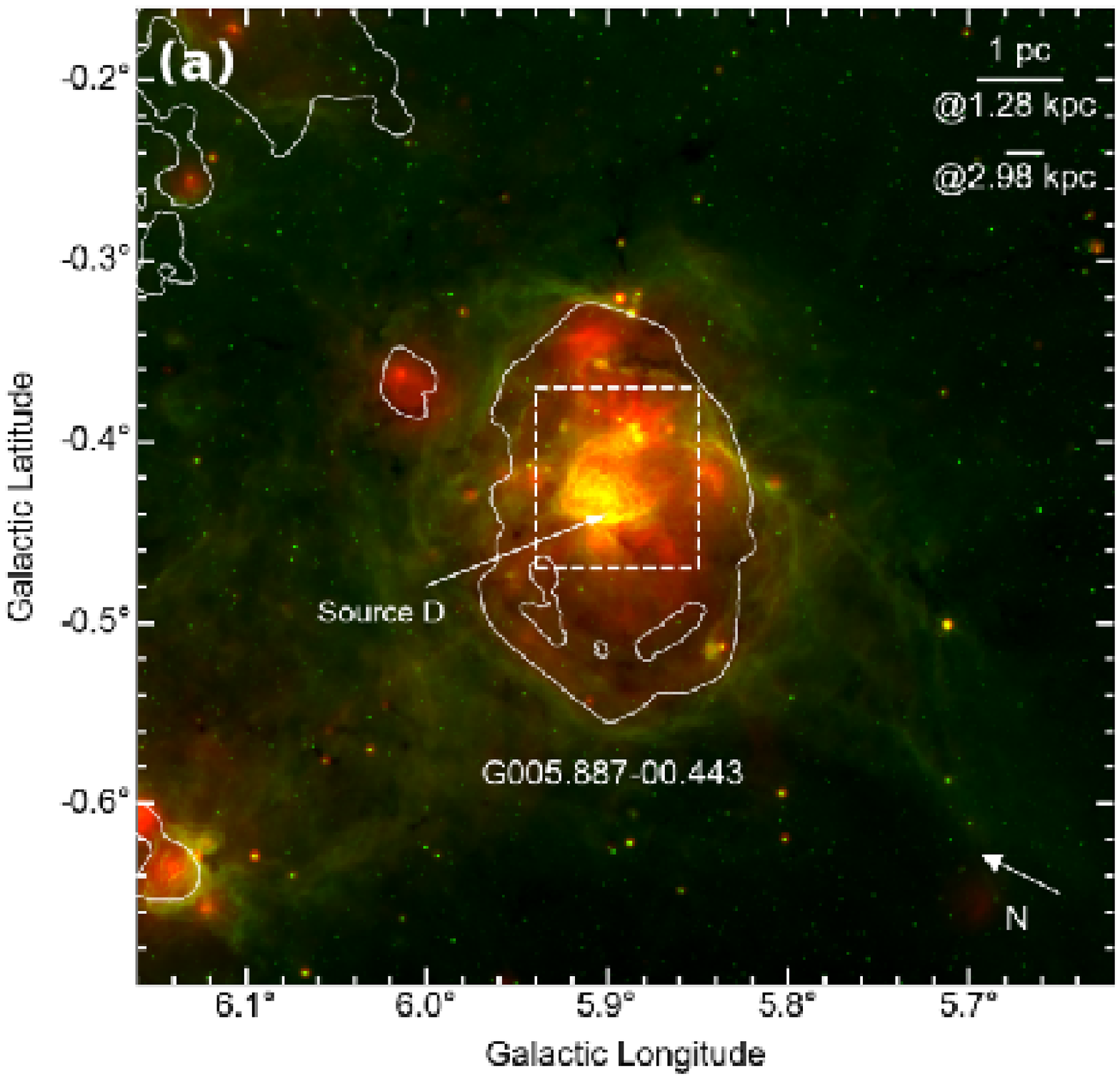}}}
   \end{center}
  \end{minipage} 
  \begin{minipage}{0.5\hsize}
   \begin{center}
    \rotatebox{0}{\resizebox{9cm}{!}{\includegraphics{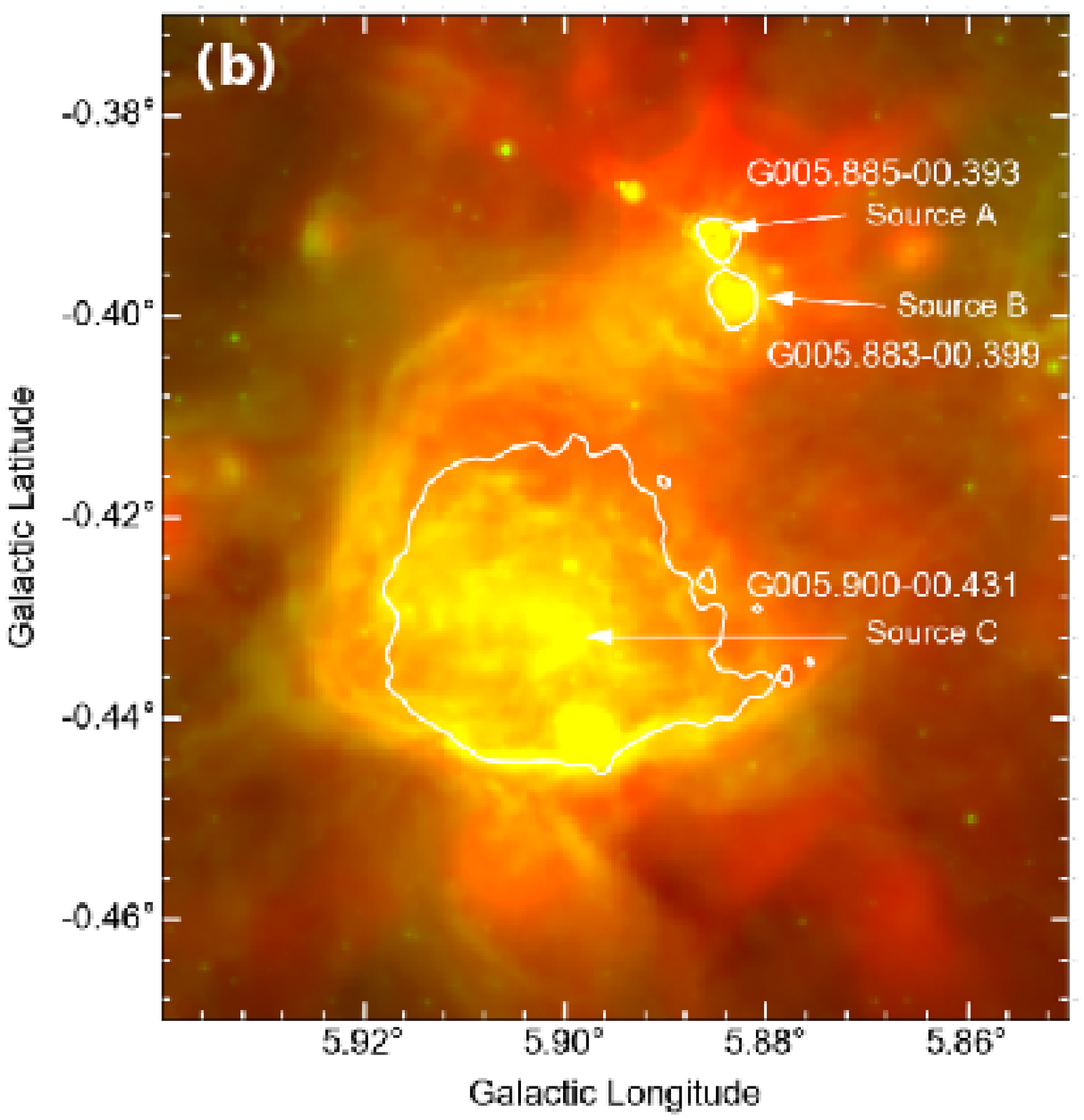}}}
   \end{center}
  \end{minipage} \\
  \end{tabular}  
  \caption{{\it Spitzer} 8$\mu$m (green) and 24$\mu$m (red) images. The dashed box in panel (a) corresponds the entire area of panel (b). The white contours show fluxes of the radio continuum emission from the VLA~20~cm data, corresponding to 2 mJy/beam in panel (a), and 10 and 16 mJy/beam respectively for the \HII\ regions of Source~C and Sources~A and B in panel (b).}
\label{fig: SpitzerCompositeImage}   
\end{figure}

W28A2 is an \HII\ complex region with bright radio continuum emission (\cite{MilneHill69}; \cite{Goudis76}), located at $\sim$~\timeform{50'} away from the supernova remnant W28. 
Figures~\ref{fig: SpitzerCompositeImage}(a) and (b) respectively show composite 8~$\mu$m and 24~$\mu$m images of this region, with contours indicating the VLA 20~cm radio continuum emission overlaid. 
Multiple \HII\ regions are identified in the {\it WISE} catalog \citep{Anderson+14}.
In the largest \HII\ region G005.887-00.443 with the radius $\sim$\timeform{0.D1}, a compact \HII\ region G005.900-00.431, and two ultra compact (UC) \HII\ regions G005.883-00.399 and G005.885-00.393 are included. 
Since W28A2 is located nearly toward the Galactic center and thus many contaminations from the line of sights are included, it is difficult to determine the distance accurately.
The distance to the UC \HII\ region G005.885-00.393 is estimated to be $\sim$2--4 kpc by kinematic studies (e.g., \cite{Acord+98}; \cite{Fish+03}) and to be 1.28 kpc \citep{Motogi+11} or 2.98 kpc \citep{Sato+14} by trigonometric parallax measurements with maser observations. 
\citet{Klaassen+06} investigated molecular gas distribution in or around G005.885-00.393 within a \timeform{100''} scale and captured the broad CO emission line with the velocity up to $\pm$50~km~s$^{-1}$.
Taking a $^{12}$CS~($J$$=$1--0) observation, \citet{Nicholas+12} found dense molecular gas toward the compact/UC \HII\ regions and an extended arm feature with the length \timeform{6'} in the northeast side of the \HII\ region.
They also detected Class I CH$_3$OH masers from the dense gas regions, suggesting the presence of shocks and outflows related to the high-mass star formation.
GeV and TeV $\gamma$ rays are detected from the W28A2 region, suggesting existence of rich gas possibly related to the supernova remnant W28 (e.g., \cite{Hanabata+14}; \cite{Hampton+16}). 
\citet{Velazquez+02} performed a large scale study of HI emission toward this area and found a strong self-absorption feature in the spectrum at $V_{\rm LSR} =$~7~km~s$^{-1}$. 

Among the \HII\ regions in W28A2, G005.885-00.393 has been widely studied as a target of high-mass protostar (known as Feldt's star) with an extremely energetic molecular outflow (e.g., \cite{Acord+98}; \cite{Feldt+03}). 
Several interferometric molecular observations discovered at least three outflows from this area (\cite{Watson+07}; \cite{Hunter+08}; \cite{Su+12}; see Figure~1 in \cite{Leurini+15}).
The spectral type of the embedded exciting star is inferred to be O or B-type zero age main sequence (ZAMS) star (e.g., O5 or earlier, based on the infrared spectral measurements: \cite{Feldt+03}; O8--O8.5, based on the Lyman continuum photons and the far-infrared luminosity: \cite{Motogi+11}), but the estimate strongly depends on the adopted distance.
An X-ray study by \citet{Hampton+16} derived the spectral type to be B5--B7, which is consistent with the lower limit given by \citet{Feldt+03}.
A short dynamical timescale of the ionized nebula ($\sim$600 yr) was inferred from the expanding ionized shell traced by the VLA measurements \citep{Acord+98} and the outflow age was estimated to be 1300--5000 yr by the measurements of the CO spectral lines \citep{Motogi+11}.
These results indicate that G005.885-00.393 is very young system.
While G005.885-00.393 is deeply investigated for a study of a star formation activity, the other \HII\ regions have been less focused yet and possible relations between the \HII\ regions are not understood. 
Furthermore, a wide molecular survey in a $\sim$\timeform{0.D3} scale beyond the \HII\ region has not been performed yet, and thus the physical associations between the massive stars and surrounding gas in the entire W28A2 region is not clear.

In this paper, we aim to investigate molecular clouds in the W28A2 region and to look for a relationship between the gas dynamics and its star formation activity, mainly focusing on the models of the triggered star formation introduced in Section \ref{sec: Triggered high-mass star formation}.
We used new $^{12}$CO and $^{13}$CO~($J$$=$1--0) data and $^{12}$CO~($J$$=$2--1) data taken by the NANTEN2 and Mopra observations with the scale of \timeform{0.D6}$\times$\timeform{0.D6}, as well as archival data set of the Mopra $^{12}$CS~($J$$=$1--0) line\footnote{http://www.physics.adelaide.edu.au/astrophysics/MopraGam/}, VLA 20~cm radio continuum emission\footnote{http://sundog.stsci.edu} and {\it Spitzer} 8 $\mu$m and 24 $\mu$m emission\footnote{https://irsa.ipac.caltech.edu/data/SPITZER/GLIMPSE/}. 
This is the first study of a probable process of high-mass star formation in the W28A2 region based on the correlation between the \HII\ regions and the surrounding gas.
Because the distance to the W28A2 region is not clear, here we adopted two distances, 1.28 kpc \citep{Motogi+11} and 2.98 kpc \citep{Sato+14}, which are results from the recent trigonometric parallax measurements, and calculated physical quantities in the two cases.
The different distance does not change our conclusion.
We also assume that each \HII\ region holds one exciting star, which are denoted as Sources A, B, C, and D for G005.885-00.393, G005.883-00.399, G005.900-00.431 and G005.887-00.443 areas, respectively. 

This paper is organized as follows. Section 2 describes the data we used in this study. 
Section~3 shows the results and Section~4 discusses possible mechanisms of the high-mass star formation in W28A2. Section~5 gives a summary of this study.

\clearpage

\section{Observations}
\label{sec:observation}

\subsection{NANTEN2 $^{12}$CO and $^{13}$CO~($J$$=$1--0), and $^{12}$CO~($J$$=$2--1) Observations}
\label{sec:nanten2coj2to1}

Observations of the $^{12}$CO and $^{13}$CO~($J$$=$1--0) lines toward the W28A2 region were made with the NANTEN2 millimeter/sub-millimeter telescope located in Atacama, Chile, in November 2011.
The frontend was a 4-K cooled Nb superconductor-insulator-superconductor mixer receiver, which provided a system noise temperature including the atmosphere $\sim$160 K in the double-side-band.
The backend was a digital spectrometer with 16384 channels at 1 GHz bandwidth and the frequency resolution was 61 kHz, which corresponds to the velocity coverage of $\sim$2600 km~s$^{-1}$ and the velocity resolution of 0.16 km~s$^{-1}$ at 115 GHz.
The pointing accuracy was achieved to be $<$~$\timeform{100''}$ by observing IRC 10216 at (R.A., Dec.)=(\timeform{09h47m57.4s}, \timeform{13D16'43.6''}) and the edge of the Sun.
The absolute intensity calibration was made with a CO observation toward $\rho-$Ophiucus at (R.A., Dec.)=(\timeform{16h19m20.9s}, \timeform{-24D22'13.0''}).
These data were smoothed to be a beam size of \timeform{200''} (HPBW) with a Gaussian function with \timeform{120''} and to be a velocity resolution of 0.5~km~s$^{-1}$ .
These corrections finally give typical rms noise levels for the $^{12}$CO and $^{13}$CO ($J$$=$1--0) lines $\sim$0.3~K~per channel and $\sim$0.5~K~per channel, respectively.
 
The NANTEN2 $^{12}$CO~($J$$=$2--1) line observation toward W28A2 was made in November 2008.
The system noise temperature including the atmosphere was $\sim$200 K at 230 GHz in the single-side-band. 
The backend was the acoustic-optical spectrometer (AOS) with 2048 channels, which corresponds to the velocity coverage of 392 km~s$^{-1}$ with the velocity resolution of 0.38 km~s$^{-1}$.
The data were smoothed to be a beam size of \timeform{100''} and to be a velocity resolution of 0.5 km~s$^{-1}$, giving a typical noise level of $\sim$0.3~K~per channel.

\subsection{Mopra $^{12}$CO and $^{13}$CO ($J$$=$1--0) Observations}
\label{sec:mopraco}

To investigate more spatially resolved gas distribution, we used the $^{12}$CO and $^{13}$CO~($J$$=$1--0) lines obtained by the Mopra 22-m millimeter telescope of the CSIRO Australia Telescope National Facility.
On-the-fly mapping observations toward W28A2 were conducted from April to July in 2016 and from April to May in 2018 as a part of the Mopra Southern Galactic Plane CO Survey (\cite{Burton+13}; \cite{Braiding+15}; \yearcite{Braiding+18}).
For the measurements of pointing accuracy, an SiO maser source VX~Sgr was used.
The typical system noise temperature was $\sim$300--800 K in the single-side-band.
We used the digital spectrometer UNSW Mopra Spectrometer (MOPS), which provides the data with the velocity resolution of 0.1 km~s$^{-1}$ and covering the velocity range $\sim$1100 km~s$^{-1}$ and $\sim$770 km~s$^{-1}$ for the $^{12}$CO and $^{13}$CO lines, respectively.
To obtain the absolute intensity, we adopted the extended beam efficiency $\eta =$ 0.55 \citep{Burton+13} for both lines. 
The original spatial resolution of the data (\timeform{36''}) is smoothed to be \timeform{45''}.
The velocity axis is smoothed to be 0.5 km~s$^{-1}$, improving the noise level up to $\sim$0.6~K~per channel and $\sim$0.3~K~per channel for the $^{12}$CO and $^{13}$CO lines, respectively.

\clearpage

\section{Results}
\label{sec:results}

\subsection{Distribution of the molecular gas}
\label{sec:DistMolGas}

\begin{figure}[h]
 \begin{center}
 \centering
  \includegraphics[width=8.0cm]{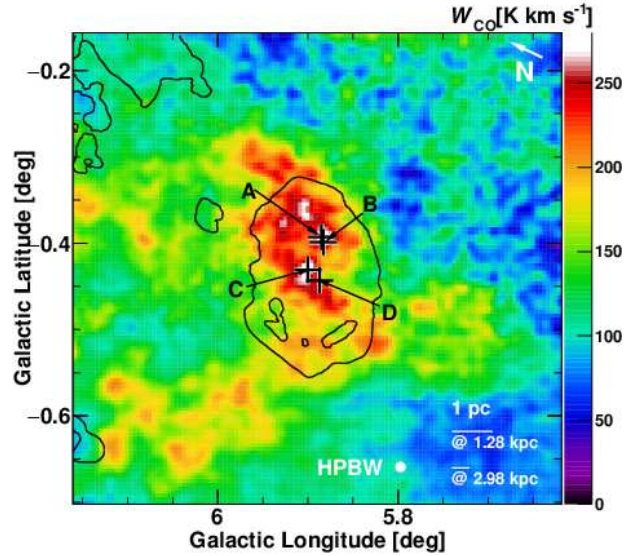}
  \end{center}
 \caption{Mopra $^{12}$CO~($J$$=$1--0) integrated intensity map with the velocity range $-10$ to 30 km~s$^{-1}$. The contours indicate an intensity of the radio continuum emission from the VLA~20~cm data drawn in Figure~\ref{fig: SpitzerCompositeImage}(a). The crosses represent positions of the exciting sources A--D.}
\label{fig: Mopra12WCO1-0}  
\end{figure}

We present the distribution of molecular clouds toward the \HII\ region W28A2, using the NANTEN2 and Mopra CO data.
Figure~\ref{fig: Mopra12WCO1-0} shows a velocity-integrated intensity map of the Mopra $^{12}$CO~($J$$=$1--0) data, with the contours of the VLA radio continuum emission.
The integrated velocity range is $-10$ to 30 km~s$^{-1}$, which covers all the velocity components found by the NANTEN2 $^{12}$CO~($J$$=$1--0) and ($J$$=$2--1) observations (Figure~\ref{fig: N2COspectra}).
We found strong emission toward the \HII\ regions and several clouds across the Galactic longitude direction in the area with $b \lesssim$ \timeform{-0.D5}.
The sources A and B are located close to the intensity peak at ($l$, $b$) $\sim$ (\timeform{5.D89}, \timeform{-0.D40}) and sources C and D are close to another intensity peak at ($l$, $b$) $\sim$ (\timeform{5.D90}, \timeform{-0.D43}).
The peak positional correspondence between the CO and radio continuum emissions suggests that these exciting sources are embedded in the rich molecular gas.
The diffuse molecular gas is widely distributed within $\sim$5~pc (@1.28 kpc) or $\sim$10~pc (@2.98 kpc) away from the exciting stars, beyond the extent of the radio continuum emission (black contour), partly showing anti-correlation with the \HII\ regions located at ($l$, $b$) $\sim$ (\timeform{6.D15}, \timeform{-0.D65}) and (\timeform{6.D12}, \timeform{-0.D25}) (see also Figure~\ref{fig: SpitzerCompositeImage}(a)).  

\begin{figure}[h]
 \begin{center}
 \centering
  \includegraphics[width=16.0cm]{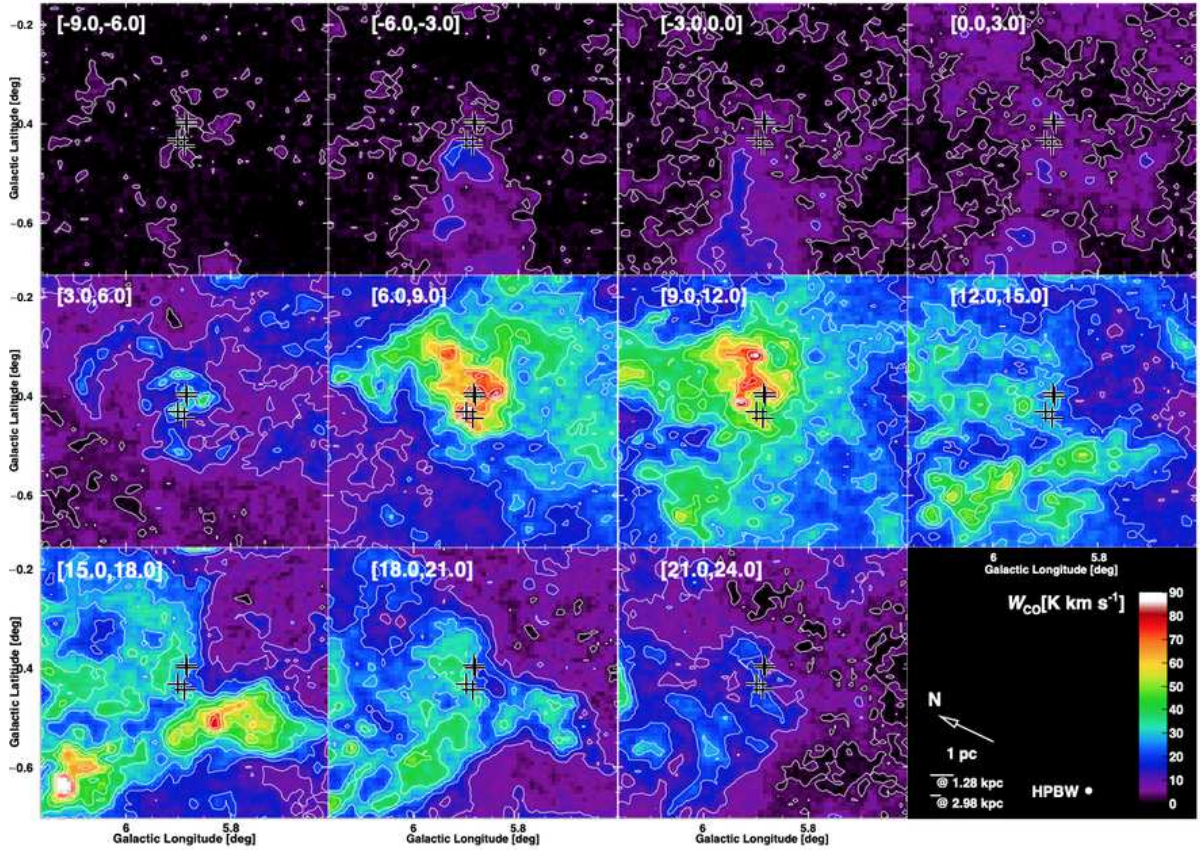}
  \end{center}
 \caption{A velocity channel map obtained with the Mopra $^{12}$CO~($J$$=$1--0) data. The integrated velocity range of each panel is 3 km~s$^{-1}$. The lowest contour level is 5$\sigma$ and the others are drawn by 20$\sigma$ step. The crosses represent positions of the exciting sources A--D.}
\label{fig: Mopra12CO10Channel}  
\end{figure}

\begin{figure}[h]
 \begin{center}
 \centering
  \includegraphics[width=15.0cm]{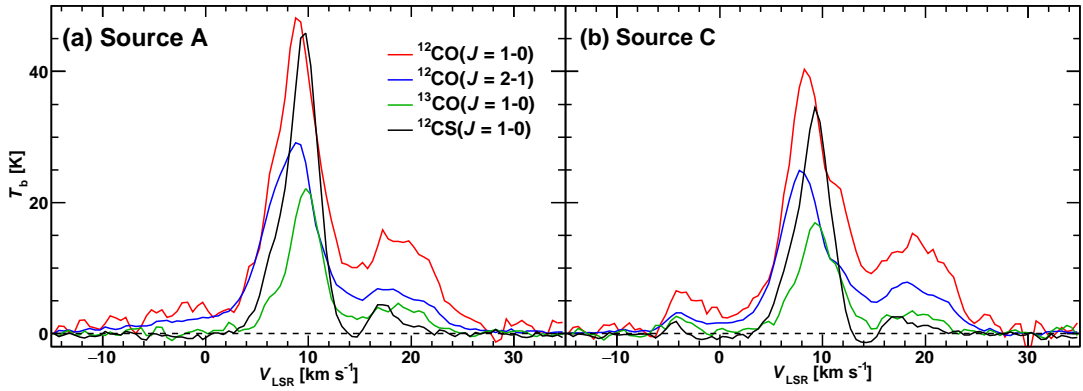}
  \end{center}
 \caption{Spectra toward Source A (a) and Source C (b): $^{12}$CO~($J$$=$1--0, $J$$=$2--1) and $^{13}$CO~($J$$=$1--0) spectra obtained by the NANTEN2 observation and $^{12}$CS~($J$$=$1--0) spectrum obtained by the Mopra observation \citep{Nicholas+12}. The $^{12}$CS~($J$$=$1--0) spectrum is scaled by a factor of 10.}
\label{fig: N2COspectra}  
\end{figure}

Figure~\ref{fig: Mopra12CO10Channel} shows a velocity channel map obtained from the Mopra $^{12}$CO~($J$$=$1--0) data.
From the morphological structure, we found three gas components separated by the velocity: low-intensity components elongated to the galactic north-south direction in the channel maps of $-6$~$< V_{\rm LSR} <$~0~km~s$^{-1}$; the high-intensity components covering the positions of the sources A--D at $V_{\rm LSR}$$\sim +9$~km~s$^{-1}$; and multiple components distinctly distributed at 12~km~s$^{-1}$ $\lesssim$ $V_{\rm LSR}$ with the highest intensity at $V_{\rm LSR}$$\sim +16$~km~s$^{-1}$.
Hereafter we denote these three molecular clouds separated by the velocity $-4$~km~s$^{-1}$ cloud, $+9$~km~s$^{-1}$ cloud and $+16$~km~s$^{-1}$ cloud, whose velocity ranges are determined by eye as $-6$ to $-1$~km~s$^{-1}$, $+4$ to $+12$~km~s$^{-1}$ and $+14$ to $+24$~km~s$^{-1}$, respectively.
These molecular gas is extending over the intermediate velocity range ($V_{\rm LSR}$$\sim -1$ to $+4$~km~s$^{-1}$ and $\sim +12$ to $+14$~km~s$^{-1}$).
Modifying the velocity range within $\sim$$\pm1$~km~s$^{-1}$ does not affect our discussion.

The $-4$~km~s$^{-1}$ cloud extended to the eastern (negative latitude) side has not been recognized in previous studies of the W28A2 region.
The positional correspondence of the CO peaks at $+9$~km~s$^{-1}$ with the exciting sources A--D suggests that these stars are embedded in the $+9$~km~s$^{-1}$ cloud.
This result is consistent with previous molecular observations (\cite{Klaassen+06}; \cite{Nicholas+12}).
The $+16$~km~s$^{-1}$ cloud has two CO peaks, one of which located in the eastern area is possibly associated another \HII\ region at ($l$, $b$) $\sim$ (\timeform{6.D15}, \timeform{-0.D65}) (see Figure~\ref{fig: SpitzerCompositeImage}(a)).
We also found a ring-like structure centered on ($l$, $b$) $\sim$ (\timeform{6.D05}\timeform{-0D35}) at $V_{\rm LSR}$$\sim$$+16$~km~s$^{-1}$ and the sources A--D are positioned at the the rim of the ring, close to one of the CO peak at ($l$, $b$) $\sim$ (\timeform{5.D83}, \timeform{-0.D52}), which is a part of the largely elongated clouds from the northeast to the southwest.

Figures~\ref{fig: N2COspectra} (a) and (b) show $^{12}$CO~($J$$=$1--0, $J$$=$2--1) and $^{13}$CO~($J$$=$1--0) spectra obtained by the NANTEN2 observations toward Sources A and C, respectively.
The three velocity components are confirmed in the spectra for Source C.
The spectra for Source A have peaks at $V_{\rm LSR}$$\sim$$9$~km~s$^{-1}$ and $\sim$$16$~km~s$^{-1}$, but the $^{12}$CO lines at $V_{\rm LSR}$$\sim$$-4$~km~s$^{-1}$ exhibit wing-like shapes rather than peaking structures.
We do not detect the significant $^{13}$CO~($J$$=$1--0) emission at $V_{\rm LSR}$$\sim$$-4$~km~s$^{-1}$ for Source~A.
For both regions, the spectral shapes with the peaks at $V_{\rm LSR}$$\sim$$9$~km~s$^{-1}$ are similar between the $^{12}$CO and $^{13}$CO ($J$$=$1--0) lines, although the $^{12}$CO~($J$$=$2--1) lines less matches the other two $J$$=$1--0 lines; the intensity of the $^{12}$CO~($J$$=$2--1) lines tends to be higher at $V_{\rm LSR}$ $\lesssim$ $9$~km~s$^{-1}$ but to be lower at $V_{\rm LSR}$ $\gtrsim$ $9$~km~s$^{-1}$.
In the same figure, we overlay the $^{12}$CS ($J$$=$1--0) line toward Sources~A~and~C, whose data were used in a previous study of the molecular clouds in the W28A2 region \citep{Nicholas+12}. 
Since the CS line has a higher critical density than CO line, the spectrum more traces emission from the high-density region, where the 
 $^{12}$CO spectrum is often saturated due to the self absorption.
For easier comparison of the spectral shape with the other $^{12}$CO lines, the intensity of the $^{12}$CS ($J$$=$1--0) line is scaled by a factor of 10.
In both regions, the peak velocities of the $^{12}$CS line are tend to be blue-shifted compared to the $^{12}$CO lines, and are almost consistent with the optically-thin line, $^{13}$CO~($J$$=$1--0) spectrum.
These results suggest a self absorption in the $^{12}$CO spectra by the dense gas in the $+9$~km~s$^{-1}$ cloud.  

\clearpage

\clearpage

\subsection{Velocity structure}
\label{sec:VelStr}

\begin{figure}[h]
 \begin{center}
 \centering
  \includegraphics[width=8.0cm]{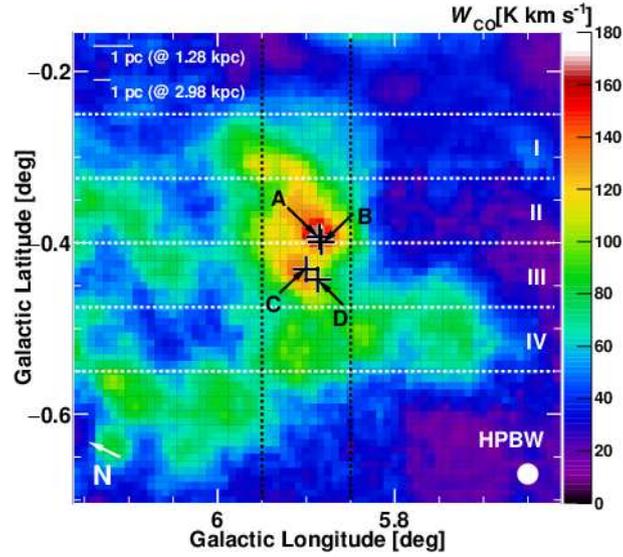}
  \end{center}
 \caption{Velocity-integrated intensity map of the NANTEN2 $^{12}$CO~($J$$=$2--1) data. The integrated velocity range is from $-$10 to $+$30 km~s$^{-1}$. The crosses represent positions of the exciting sources A--D. The white dotted lines indicate the integrated latitude range of the position-velocity diagram shown in Figure~\ref{fig: N212CO21LV}. The black dotted lines indicate the integrated longitude range of the position-velocity diagram shown in Figure~\ref{fig: N212CO21to10VB}.} 
\label{fig: N212WCO21}  
\end{figure}

\begin{figure}[h]
 \begin{center}
 \centering
  \includegraphics[width=16.0cm]{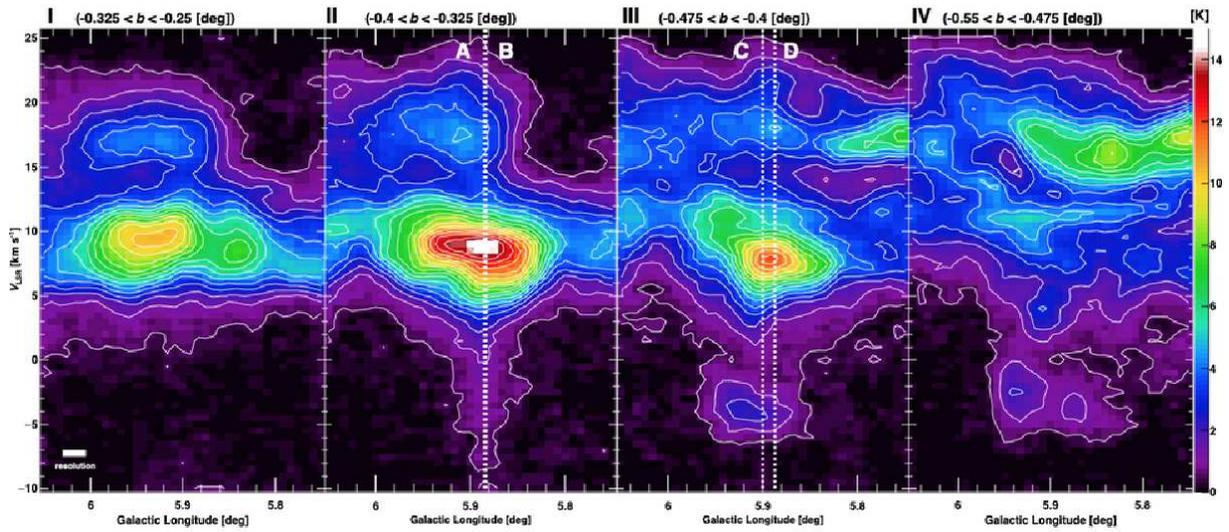}
  \end{center}
 \caption{Longitude-velocity diagrams of the NANTEN2 $^{12}$CO~($J$$=$2--1) data for the regions I--IV shown in Figure~\ref{fig: N212WCO21}. The white dotted lines indicate positions of the exciting sources A--D. The lowest contour corresponds to 5$\sigma$ level and the others are drawn by 6$\sigma$ interval.}
\label{fig: N212CO21LV}  
\end{figure}

Figure~\ref{fig: N212WCO21} shows a velocity-integrated ($-10$~km~s$^{-1}$ $<$ $V_{\rm LSR}$ $<$ $+$30 km~s$^{-1}$) intensity map of the NANTEN2 $^{12}$CO~($J$$=$2--1) data.
We separated the map into four regions I--IV for the different latitude range and made longitude-velocity diagrams for each region as shown in Figure~\ref{fig: N212CO21LV}.
In the regions~I--III, the highest CO emissions are found at $V_{\rm LSR} \sim$~9 km~s$^{-1}$, while the region IV has the CO intensity peak at $V_{\rm LSR} \sim$~16 km~s$^{-1}$.
In the region III, where Sources C and D are included, the low-intensity gas for the $-4$ km~s$^{-1}$ cloud is confirmed and it is connected to the $+$9 km~s$^{-1}$ cloud with a bridging feature at $V_{\rm LSR} \sim$~1 km~s$^{-1}$.
The region~II including Sources A and B also shows a presence of the low-intensity gas at $V_{\rm LSR} \sim$~$-$4 km~s$^{-1}$, and it is connected to the $+9$ km~s$^{-1}$ cloud along the direction of these sources.
This wing-like strucutre is also found in the opposite side at $V_{\rm LSR} \sim$~13~km~s$^{-1}$, just coinciding with the direction of Sources A and B. 
The region~IV shows a distinct cloud at $V_{\rm LSR} \sim$~$-$4~km~s$^{-1}$ with a bridging feature relatively extended in \timeform{5.D85} $\lesssim l \lesssim$ \timeform{5.D97} at $V_{\rm LSR} \sim$~1 km~s$^{-1}$.
We do not find the significant emission from the $-4$ km~s$^{-1}$ cloud and the associated bridging feature in the region~I.
In addition to the wing-like structure at $V_{\rm LSR} \sim$~13~km~s$^{-1}$ in the region~II, the intermediate velocity components between the $+$9 km~s$^{-1}$ and 16 km~s$^{-1}$ clouds are found in all regions, but they are extensively distributed and have more complicated velocity structure especially for the regions~III and~IV. 

\begin{figure}[h]
 \begin{center}
 \centering
  \includegraphics[width=10.0cm]{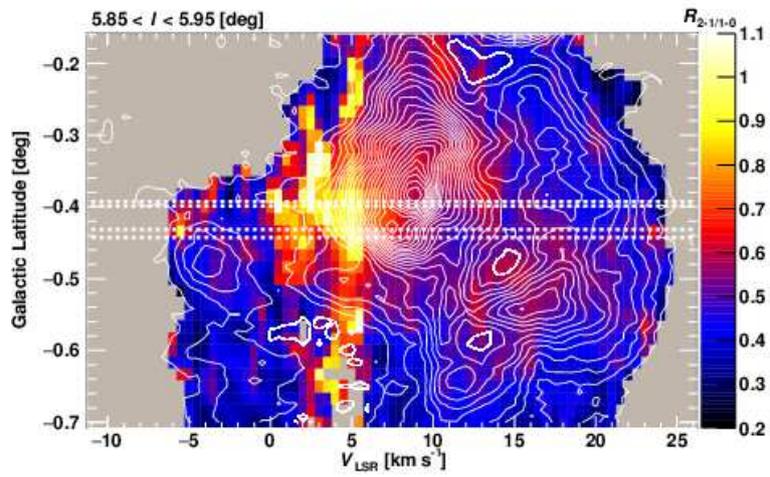}
  \end{center}
 \caption{Latitude-velocity diagram of the 2--1 to 1--0 intensity ratio (image) of the NANTEN2 $^{12}$CO data. The integrated longitude range is from \timeform{5.D85} to \timeform{5.D95}, which is shown by the black dotted lines in Figure~\ref{fig: N212WCO21}. The white contours indicate the NANTEN2 $^{12}$CO~($J$$=$2--1) data, drawn in 6$\sigma$ step from 5$\sigma$ intensity level. The thick contours show the depletion area compared to the surrounding area. The white dotted lines indicate the positions of the exciting sources A--D from top to bottom.}
\label{fig: N212CO21to10VB}  
\end{figure}

Figure~\ref{fig: N212CO21to10VB} shows a latitude-velocity diagram of the NANTEN2 $^{12}$CO~($J$$=$2--1) to ($J$$=$1--0) intensity ratio for the integrated longitude range from \timeform{5.D85} to \timeform{5.D95} represented by the black dotted lines in Figure~\ref{fig: N212WCO21}. 
For comparison, contours from the $^{12}$CO~($J$$=$2--1) data are superposed.
We note that the intensity ratio map using the Mopra $^{12}$CO data also gives the similar latitude-velocity diagram.
Overall, the ratio becomes lower as far away from the exciting sources A--D shown by the dotted lines, except for the local high intensity ratio at ($V_{\rm LSR}$, $b$) $=$ ($\sim$5 km~s$^{-1}$, \timeform{-0.D25}) and ($\sim$5 km~s$^{-1}$, \timeform{-0.D65}). 
The intensity ratio is as high as $\sim$~1.0 particularly at $V_{\rm LSR} \sim$~0--5~km~s$^{-1}$.
At $V_{\rm LSR} \sim$ $9$~km~s$^{-1}$, where the physical correlation between the gas and embedded stars are expected, the intensity ratio is down to $\sim$0.6.
The velocity with the highest intensity ratio is not consistent with that of the CO emission peak at $V_{\rm LSR} \sim$~$9$~km~s$^{-1}$. 
This is possibly due to self absorption in the rich gas of the 9 km~s$^{-1}$ cloud as discussed in Section~\ref{sec: PhysicalAssociationMCandStars}.
The local CO intensity peak at $V_{\rm LSR} \sim$ $16$~km~s$^{-1}$, which corresponds to a main cloud in the $+16$ km~s$^{-1}$ cloud, shows the intensity ratio $\sim$0.6.
At $V_{\rm LSR} \sim$ $-4$~km~s$^{-1}$, the intensity ratio has tend to be higher ($\sim$0.6--0.9) toward the direction for the exciting stars.

\clearpage

\subsection{Comparison with the radio continuum emission and the infrared emission}
\label{sec: GasVLA8um}

\begin{figure}[h]
 \begin{tabular}{cc}
  \begin{minipage}{0.5\hsize}
   \begin{center}
    \rotatebox{0}{\resizebox{8cm}{!}{\includegraphics{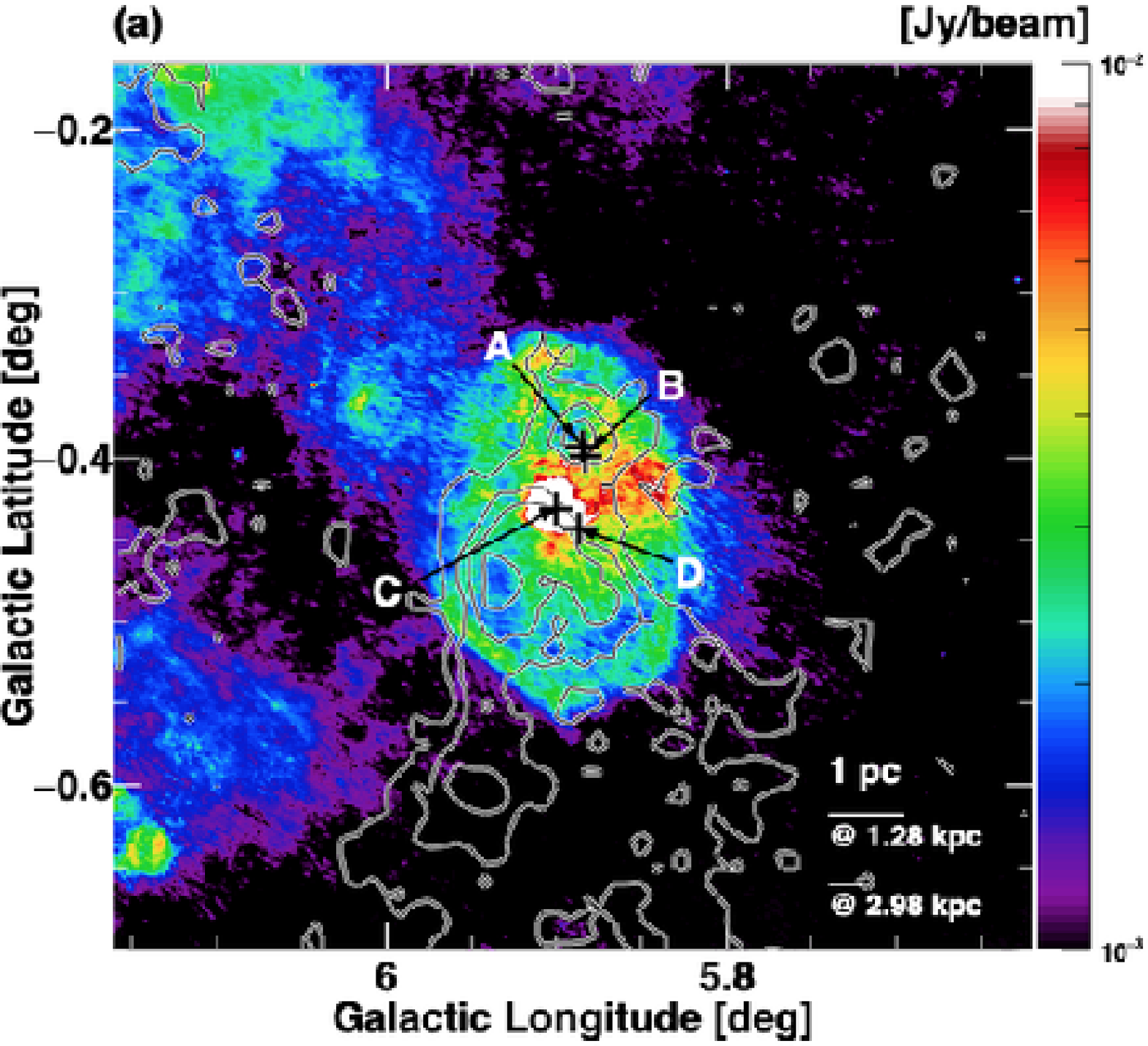}}}
   \end{center}
  \end{minipage} 
  \begin{minipage}{0.5\hsize}
   \begin{center}
    \rotatebox{0}{\resizebox{8cm}{!}{\includegraphics{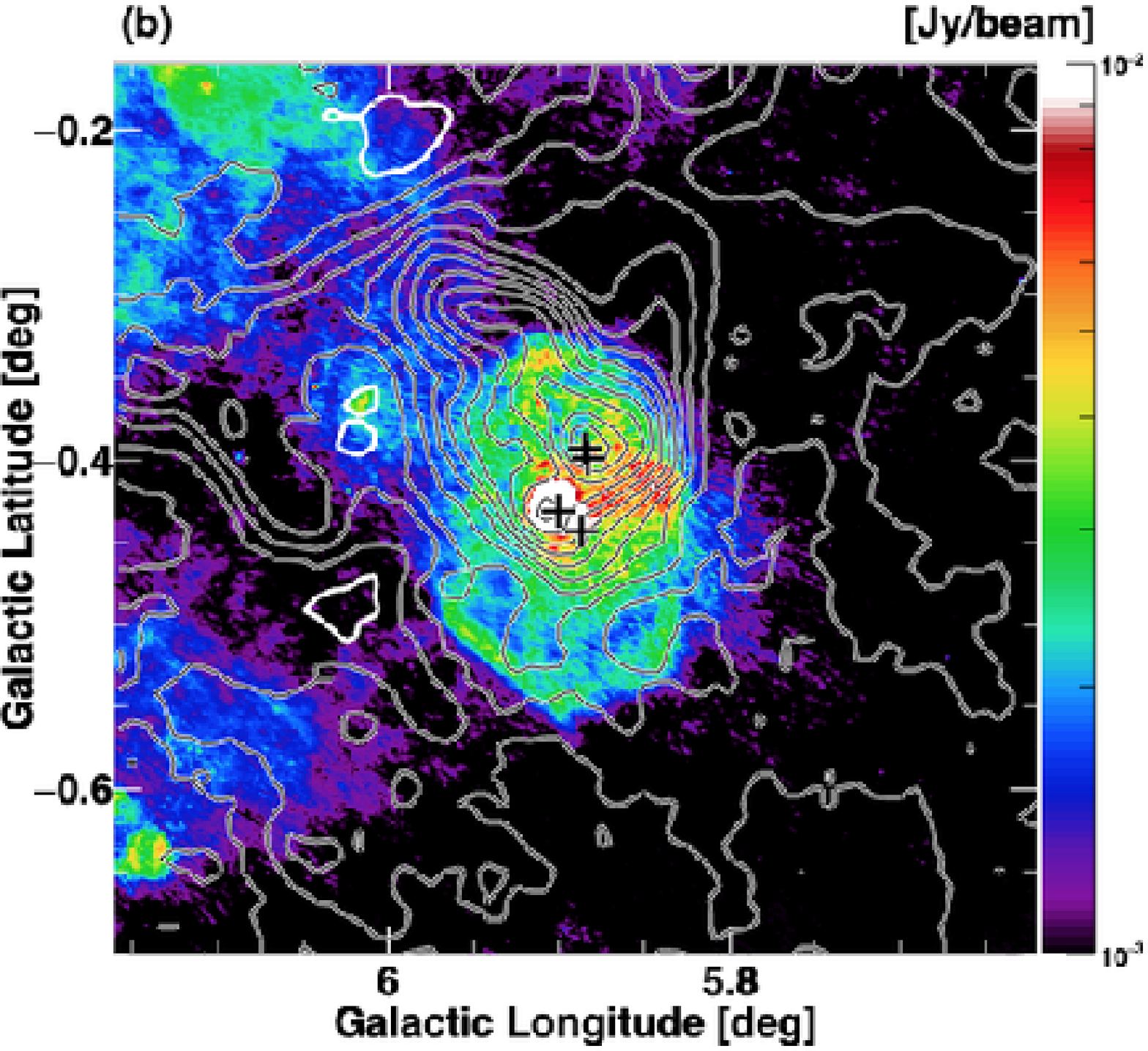}}}
   \end{center}
  \end{minipage} \\
    \begin{minipage}{0.5\hsize}
   \begin{center}
    \rotatebox{0}{\resizebox{8cm}{!}{\includegraphics{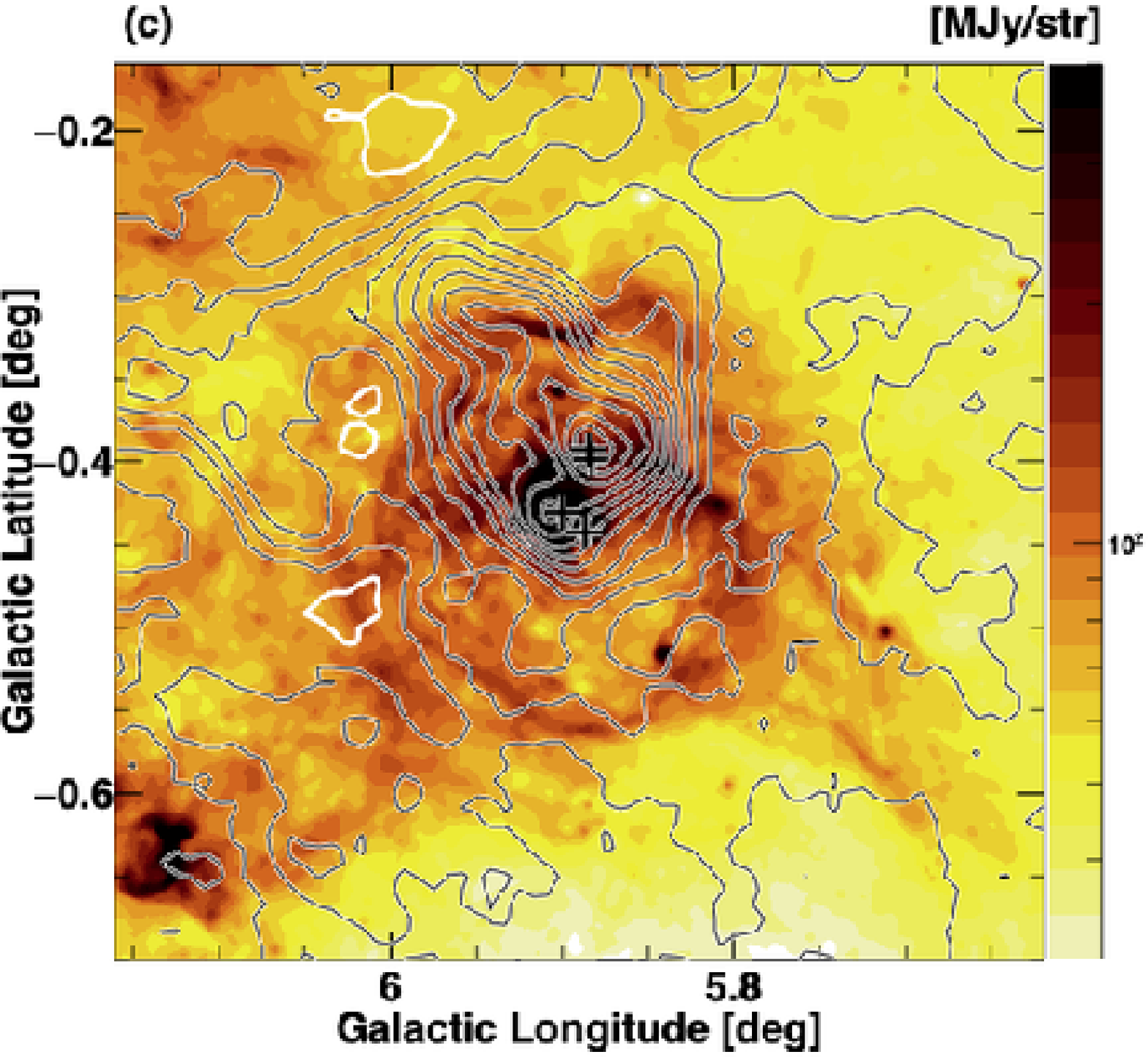}}}
   \end{center}
  \end{minipage} 
  \begin{minipage}{0.5\hsize}
   \begin{center}
    \rotatebox{0}{\resizebox{8cm}{!}{\includegraphics{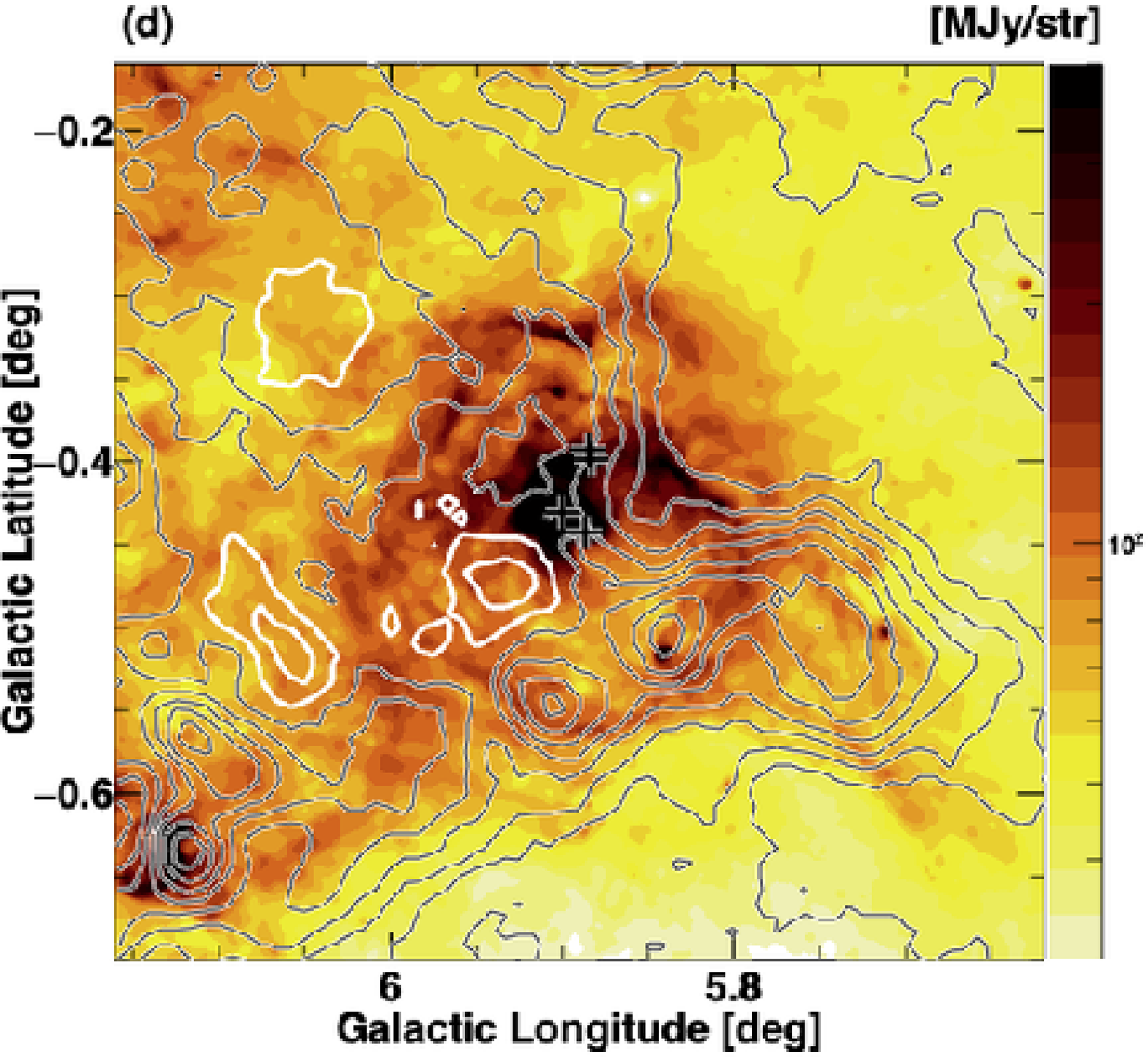}}}
   \end{center}
  \end{minipage} 
  \end{tabular}  
  \caption{Molecular gas distribution of the NANTEN2 $^{12}$CO~($J$$=$2--1) data (gray contours) in comparison with the VLA 20 cm radio continuum emission and {\it Spitzer} 8$\mu$m emission (image). (a) $-4$ km~s$^{-1}$ cloud vs. VLA 20 cm; (b) $+9$ km~s$^{-1}$ cloud vs. VLA 20 cm; (c) $+9$ km~s$^{-1}$ cloud vs. {\it Spitzer} 8$\mu$m; (d) $+16$ km~s$^{-1}$ cloud vs. {\it Spitzer} 8$\mu$m. The contours start from 5$\sigma$, 15$\sigma$ and 10$\sigma$, and are drawn in 6$\sigma$, 12$\sigma$ and 11$\sigma$ steps for (a), (b-c) and (d), respectively. The white contours in panels (b)--(d) indicate depression areas compared to the surrounding area. The crosses represent positions of the exciting sources A--D.}
\label{fig: N212CO21compVLA20cmSpitzer8um}   
\end{figure}

In order to investigate physical correlations between the gas and the exciting stars, we made comparisons with spatial distributions of the radio continuum emission possibly related to the UV photons radiated from the exciting stars and the 8 $\mu$m infrared emission mostly from the polycyclic aromatic hydrocarbon, which are formed by photodissociation due to the UV radiation from the exciting stars. 
Here we used the NANTEN2 $^{12}$CO~($J$$=$2--1) data, which have higher sensitivity than the Mopra $^{12}$CO~($J$$=$1--0) data. 
Figures~\ref{fig: N212CO21compVLA20cmSpitzer8um}~(a) and (b) show distributions of the radio continuum emission obtained by VLA (image), compared to the molecular gas distributions of the NANTEN2 $^{12}$CO~($J$$=$2--1) data for the $-4$ km~s$^{-1}$ cloud and the $+9$ km~s$^{-1}$ cloud, respectively (gray contours).
The white contours indicate depression areas in CO compared to the surrounding area. 
The CO intensity is enhanced at $(l,b) =$ (\timeform{5.D94}, \timeform{-0.D47}) for the $-4$ km~s$^{-1}$ cloud and depressed at $(l,b) =$ (\timeform{6.D02}, \timeform{-0.D38}) for the $+9$ km~s$^{-1}$ cloud, while conversely the radio continuum emission is depressed and enhanced, respectively.
These features suggest that UV photons radiated from the massive stars are extended to the low-density medium away from the regions with rich gas.
Similarly, in Figures~\ref{fig: N212CO21compVLA20cmSpitzer8um}~(c) and (d), the spatial distributions of the {\it Spitzer} 8~$\mu$m infrared emission are compared to the NANTEN2 $^{12}$CO~($J$$=$2--1) data for the $+9$~km~s$^{-1}$ and $+16$~km~s$^{-1}$ clouds, respectively.
The 8 $\mu$m distribution is similar to the molecular gas distribution and delineates the outer boundary of the radio continuum emission, implying that the photodissociation is proceeding at the surface of the molecular clouds.
We found spatial correspondences at $(l,b) =$ (\timeform{5.D92}, \timeform{-0.D31}) for the $+9$ km~s$^{-1}$ cloud and $(l,b) =$ (\timeform{5.D75}, \timeform{-0.D5}) for the $+16$ km~s$^{-1}$ cloud.
These results suggest that physical association of the molecular clouds of all three velocity components with the exciting sources.

\clearpage

\section{Discussion}
\label{sec:Discussion}

The results obtained in Section~\ref{sec:results} show the molecular gas properties in the W28A2 region as follows. 

\begin{itemize}

\item The molecular clouds are extended in the scale of $\sim$5--10~pc from the center of \HII\ region, separated by the velocity into three components with the CO peaks at $V_{\rm LSR} \sim$ $-4$~km~s$^{-1}$, $+9$~km~s$^{-1}$ and $+16$~km~s$^{-1}$. 
The exciting stars are embedded in gas-rich regions of the $+9$~km~s$^{-1}$ cloud, possibly forming the \HII\ complex.

\item The position-velocity diagram of the CO data shows bridging features connecting the three clouds. The wing-like structure crossing the intermediate velocity range coincides with the directions of the exciting sources A and B.
 
\item Overall, the $^{12}$CO 2--1 to 1--0 intensity ratio in the latitude-velocity diagram becomes lower as far away from the exciting sources. The high intensity ratio ($\sim$0.8--1.0) is found in the $-4$~km~s$^{-1}$ and $+9$~km~s$^{-1}$ clouds, but the velocity at the peak intensity ratio in the $+9$~km~s$^{-1}$ cloud is not consistent with the CO peak at $V_{\rm LSR} \sim$ $9$~km~s$^{-1}$. 
The intensity ratio of the $+16$~km~s$^{-1}$ cloud is relatively lower ($\lesssim$ 0.6).

\item Comparisons of the gas distributions with the radio continuum emission and the 8 $\mu$m infrared emission show spatial coincidence/anti-coincidence, suggesting physical associations between the gas and ionizing photons radiated from the exciting stars. 

\end{itemize}

\subsection{Spectral type of the exciting stars and the molecular gas mass}
\label{sec: SpectralTypeGasMass}

Assuming that each \HII\ region is formed by one exciting source, we estimated their spectral types by using the number of Lyman continuum photons ($N_{\rm Ly}$), which is derived from the equation below \citep{SimpsonRubin90},

\begin{eqnarray}
N_{\rm Ly} \simeq \frac{5.59 \times 10^{48}}{(1+f_{\rm i})} \times T_{\rm e}^{-0.45} \times \left(\frac{\nu}{\nu_{5}}\right)^{0.1}S_{\nu} D^{2}, 
\label{eq:} 
\end{eqnarray}
where $\nu$ is the frequency in GHz ($\nu_{5} =$ 5.0 GHz) and $D$ is the distance to the source. 
$f_{\rm i}$ ($\equiv {\rm He^{+}}/({\rm H^{+}+He^{+}})$ $\simeq$ 0.65) is the helium fraction of recombination photons to excited states.
$S_{\nu}$ is the flux density of each \HII\ region, whose area is determined by a threshold of the radio continuum emission shown in Figure~\ref{fig: SpitzerCompositeImage}. 
This boundary is adjusted not to overlap the different \HII\ regions and to fall within the approximate circular regions determined in the {\it WISE} catalog \citep{Anderson+14}.
In estimating the flux of Source D, we masked the regions of Sources A--C to remove their contributions and interpolated with the average flux of Source D.
The electron temperature $T_{\rm e}$ is assumed to be 6700~K, which is obtained from H$\alpha$ recombination line near the W28A2 region \citep{Downes+80}.  
Table~\ref{table: SpectralTypeExStars} shows the derived $S_{\nu}$ and $N_{\rm Ly}$ at $\nu =$ 1.4 GHz and inferred spectral types of each ionizing star under assumptions of the same distance for these \HII\ regions $D =$ 1.28 kpc \citep{Motogi+11} and 2.98 kpc \citep{Sato+14}.

The spectral type of Source A corresponding to the position of Fieldt's star is derived to be B0.5, which is consistent with previous studies if we take into account the uncertainty arising from the different estimates with an assumption of the distance (e.g., \cite{Feldt+03}; \cite{Motogi+11}; \cite{Sato+14}). 
The Source C whose spectral type is derived to be O7.5--9.5 coincides with positions of an X-ray protostar \citep{Hampton+16} and is consistent with their estimates for a late O-type star.
Among the four sources, Source D having the largest $N_{\rm Ly}$ shows the most massive stars with the spectral type O6--8.

\begin{table}[h]
  \caption{Properties of the \HII\ regions and expected spectral types of their exciting sources in W28A2} 
 \begin{center}
 \label{table: SpectralTypeExStars}
   \begin{tabular}{ccccc} \hline\hline
   \makebox[5em][c]{Exciting star} &
   \makebox[5em][c]{\HII\ region} &
   \makebox[5em][c]{$S_{\rm \nu}$ (1.4 GHz)} &
   \makebox[5em][c]{log$N_{\rm Ly}$ } &
   \makebox[5em][c]{Spectral type} \\ 
    &  & [Jy] & [photons s$^{-1}$] & (ZAMS) \\ 
    &  & & @1.28 kpc / @2.98 kpc & @1.28 kpc / @2.98 kpc \\ 
    & (1) & & & (2) \\ \hline
   Source A & G005.885--00.393 & 0.29 & 46.43 / 47.17 & $\sim$B0.5 /$\sim$B0.5 \\
   Source B & G005.883--00.399 & 0.20 & 46.27 / 47.00 & $\sim$B0.5 /$\sim$B0.5 \\
   Source C & G005.900--00.431 & 6.09 & 47.75 / 48.49 & $\sim$O9.5 / $\sim$O7.5 \\
   Source D & G005.887--00.443 & 25.71 & 48.38 / 49.12 & $\sim$O8 / $\sim$O6 \\ \hline
   \multicolumn{5}{l}{\scriptsize{*(1) {\it WISE} catalog name \citep{Anderson+14} (2) Spectral type (ZAMS) \citep{Panagia73}}}\\ 
   \end{tabular}
  \end{center}
\end{table}

The molecular gas masses of the $-4$ km~s$^{-1}$, $+9$ km~s$^{-1}$ and $+16$ km~s$^{-1}$ clouds are estimated from the NANTEN2 $^{13}$CO~($J$$=$1--0) data: using the optical depth of the $^{13}$CO~($J$$=$1--0) lines derived from an assumption of the local thermodynamic equilibrium (LTE), we calculated the column density for the $^{12}$CO by adopting the $^{13}$CO/$^{12}$CO abundance ratio 7.1$\times$10$^5$ \citep{Frerking+82}. 
Relevant equations to calculate the gas mass is summarized in \citet{Nishimura+15}.
We simply derived the gas mass for the entire region investigated in this study.
The obtained peak column density and gas masses for $D =$ 1.28 kpc and 2.98 kpc are summarized in Table~\ref{table: gas masses}.

\citet{Nicholas+12} measured the molecular gas column density using the CS ($J$$=$1--0) line and estimated the gas mass around the four exciting sources A--D within the scale of $\sim$\timeform{10'}.
The obtained $N_{\rm H2}$ was 2.9$\times$10$^{23}$ cm$^{-2}$, which is much larger than that of our estimate even compared to the total value of the three velocity components ($N_{\rm H2}$ $\sim$ 7.3$\times$10$^{22}$ cm$^{-2}$).
The larger $N_{\rm H2}$ obtained in \citet{Nicholas+12} is probably due to a high-density gas tracer CS line, which has a higher critical density than CO line. 
The spatial resolution of the CS ($J$$=$1--0) study is $\sim$\timeform{1'} \citep{Nicholas+12} and thus the beam dilution effect is more significant for the NANTEN2 $^{13}$CO~($J$$=$1--0) data, resulting in the smaller $N_{\rm H2}$ in our estimate.
If we assume the same distance adopted in \citet{Nicholas+12} ($D =$ 2 kpc), the molecular gas mass with our CO data for the region adopted in their study (see Figure 2 in \cite{Nicholas+12}) is derived to be 2.3$\times$10$^{4}$ $\Msun$, which is consistent within a factor of~2.

\begin{table}[h]
 \caption{Molecular gas column density and gas mass for the three velocity clouds} 
 \label{table: gas masses}
  \begin{center}
   \begin{tabular}{cccc} \hline\hline
   \makebox[10em][c]{Cloud} & \makebox[7em][c]{$N_{\rm H2}$ (peak)} & \makebox[7em][c]{Molecular gas mass} \\ 
   & [10$^{22}$ cm$^{-2}$] & [$M_{\odot}$] \\ 
      &  & @1.28 kpc / @2.98 kpc \\ \hline
   $-4$ km s$^{-1}$ cloud & 0.5 & 2.5$\times$10$^{3}$ / 1.3$\times$10$^{4}$ \\
   $+9$ km s$^{-1}$ cloud & 4.0 & 3.5$\times$10$^{4}$ / 1.9$\times$10$^{5}$ \\
   $+16$ km s$^{-1}$ cloud & 2.8 & 2.7$\times$10$^{4}$ / 1.4$\times$10$^{5}$ \\ \hline
   \end{tabular}
  \end{center}
\end{table}

\subsection{Physical association of the molecular clouds and the exciting stars}
\label{sec: PhysicalAssociationMCandStars}

\begin{figure}[h]
 \begin{center}
 \centering
  \includegraphics[width=7.0cm]{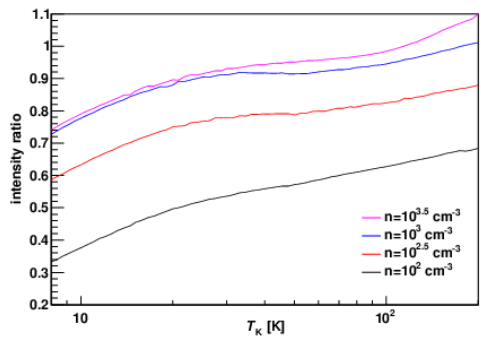}
  \end{center}
  \caption{Correlations between the kinetic temperature ($T_{\rm k}$) and the $^{12}$CO 2--1/1--0 intensity ratio for the several molecular gas densities ($n$).}
\label{fig: RatioVsTk}  
\end{figure}

In Figure~\ref{fig: N212CO21to10VB}, we found the relatively high $^{12}$CO 2--1/1--0 intensity ratio toward the exciting stars at $V_{\rm LSR} \sim$0--5~km~s$^{-1}$ and $\sim$$-4$~km~s$^{-1}$.  
Using the NANTEN2 $^{12}$CO~($J$$=$1--0, $J$$=$2--1) and $^{13}$CO~($J$$=$1--0) lines, we performed a Large Velocity Gradient (LVG) analysis \citep{GoldreichKwan74} under assumptions of the abundance ratios, [$^{12}$CO]/[H$_2$] = 10$^{-4}$ (e.g., \cite{Frerking+82}; \cite{Leung+84}) and [$^{12}$C]/[$^{13}$C] $=$ 77 \citep{WilsonRood94}, and a typical velocity gradient estimated from the CO data $dv/dr =$~0.5.
We confirmed variations of $dv/dr$ from 0.1 to 1.0 do not change the physical quantities for the target positions significantly.
The intensity ratio as a function of the kinematic temperature ($T_{\rm k}$) for the several molecular gas density is shown in Figure~\ref{fig: RatioVsTk}.
The ratio toward the direction of the exciting stars for the $-4$~km~s$^{-1}$ cloud is up to $\sim$0.9 (see  Figure~\ref{fig: N212CO21to10VB}).
From the LVG analysis, the molecular gas density for the $-4$~km~s$^{-1}$ cloud is derived to be $\sim$1$\times$10$^3$ cm~$^{-3}$. 
Thus the gas temperature at $V_{\rm LSR} \sim$ $-4$~km~s$^{-1}$ estimated from Figure~\ref{fig: RatioVsTk} is $\sim$20~K, indicating that the $-4$~km~s$^{-1}$ cloud is heated by the UV radiation from the massive stars.

For the $+9$ km~s$^{-1}$ cloud, the intensity ratio in the blueshift side (at $V_{\rm LSR} \sim$5~km~s$^{-1}$) is as large as $\sim$1.0, while the ratio in the redshift side is only up to $\sim$0.6.   
The discrepancy is expected from the different spectral structure between the $^{12}$CO~($J$$=$1--0) and ($J$$=$2--1) lines (see Figure~\ref{fig: N2COspectra})
and does not change significantly even if we apply the Mopra $^{12}$CO~($J$$=$1--0) data instead of the NANTEN2 $^{12}$CO~($J$$=$1--0) data.
A probable effect giving the lower intensity ratio in the redshift side is self absorption by the high-density gas in the near side of the $+9$~km~s$^{-1}$ cloud.
It can be also suggested from the $^{12}$CS~($J$$=$1--0) spectrum as described in Section~\ref{sec:DistMolGas}.
We derived the optical depth of the $^{13}$CO~($J$$=$1--0) line through the molecular gas mass estimate (Section~\ref{sec: SpectralTypeGasMass}). 
It is found to be 0.1--0.5 at 6~$\lesssim$ $V_{\rm LSR}$ $\lesssim$~12~km~s$^{-1}$ but is less than 0.1 at 6~km~s$^{-1}$~$\lesssim$ $V_{\rm LSR}$, suggesting that the self absorption is not significant at 6~km~s$^{-1}$~$\lesssim$ $V_{\rm LSR}$.
Applying the intensity ratio $\sim$1.0 at 6~km~s$^{-1}$~$\lesssim$ $V_{\rm LSR}$, the gas density derived by the LVG analysis is $\sim$3$\times$10$^3$ cm~$^{-3}$, and thus the gas temperature inferred from Figure~\ref{fig: RatioVsTk} is $\sim$~100~K, suggesting a strong effect by the UV radiation from the exciting stars. 
According to \citet{Velazquez+02}, the W28 region has possible large amount of clod HI in front of the \HII\ complex.
These features are also found in the $^{12}$CO~($J$$=$3--2) spectrum toward Source C as shown in Figure~3 in \citet{Klaassen+06}.
The effects of optical depths are also discussed in \citet{WoodChurchwell89}.

The intensity ratio of the $+16$~km~s$^{-1}$ cloud overall tends to be lower as far away toward the negative latitude direction from the exciting stars (see Figure~\ref{fig: N212CO21to10VB}).
This result may indicate that the $+16$~km~s$^{-1}$ cloud is also affected by the UV radiation from Sources A--D.
The spatial coincidence between the gas and 8~$\mu$m infrared emission found in Section~\ref{sec: GasVLA8um} supports this indication. 
However, the intensity ratio ($\sim$0.6) is lower than the other two clouds, implying that the $+16$~km~s$^{-1}$ is located most far away from the exciting stars. 

W28A2 is located toward a near direction of the Galactic center and thus multiple gas components across different spiral arms might be observed on the same line of sight.
The $-4$ km~s$^{-1}$ cloud is relatively isolated and any continuous gas distribution with the same velocity are not found in the CO data.
This result suggests that the $-4$ km~s$^{-1}$ cloud is not a single component included in another spiral arm.
The $+9$ km~s$^{-1}$ and $+16$ km~s$^{-1}$ clouds mainly cover the velocity range belonging to the Scutum or Norma arms \citep{Reid+16}.
Although it is not clear that these clouds belong to either arm, we conclude that they are located in a proximal space in the same arm since they exhibit physical associations with the \HII\ region described in Sections~\ref{sec: GasVLA8um} and \ref{sec: TrigMechanisum}.

\subsection{Molecular outflow}
\label{sec:MolecularOutflow}

\begin{figure}[h]
 \begin{center}
 \centering
  \includegraphics[width=7.0cm]{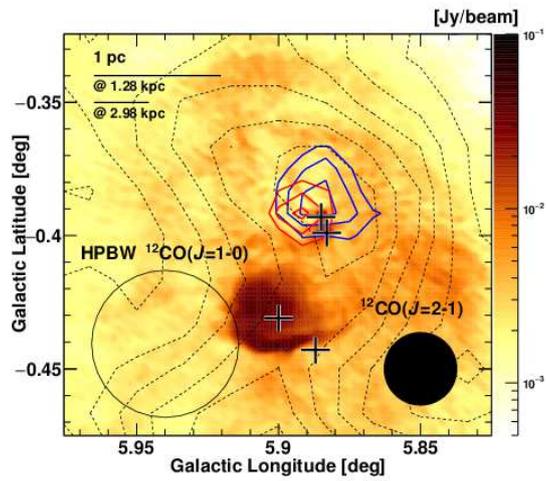}
  \end{center}
  \caption{The velocity-integrated maps for the wing components using the NANTEN2 $^{12}$CO~($J$$=$2--1) data (contours) superposed on the VLA 20~cm radio continuum emission (image). The integrated velocity ranges for the wing components are $-15$ km~s$^{-1}$ to $+3$ km~s$^{-1}$ (blue) and $+15$ km~s$^{-1}$ to $+30$ km~s$^{-1}$ (red). The contours with the dotted lines indicate the intermediate velocity component ($-1$ km~s$^{-1}$ $< V_{\rm LSR}$ $< +4$ km~s$^{-1}$) with the NANTEN2 $^{12}$CO~($J$$=$1--0) data, drawn by 2 $\sigma$ step from the 5 $\sigma$ level. The beam sizes of each CO line are represented. The crosses represent positions of the exciting sources A--D.} 
\label{fig: OutflowMap}  
\end{figure}

As described in Section~\ref{sec:W28A2}, the Feldt's star, which forms one of an UC \HII\ region in the W28A2 complex, having an extraordinary bipolar molecular outflow, has been studied extensively (e.g. \cite{Feldt+03}; \cite{Leurini+15}). 
This massive protostar corresponds to Source A, around which we find wing-like structures in the CO spectrum shown in Figure~\ref{fig: N2COspectra} and the position-velocity diagram for the region~II in Figure~\ref{fig: N212CO21LV}.  
Its broad line emission covers the velocity range from $V_{\rm LSR}$ $\sim$ $-15$ km~s$^{-1}$ to $\sim$ $+30$ km~s$^{-1}$, centered on the systemic velocity at $V_{\rm LSR} \sim$ $+9$~km~s$^{-1}$.
Using the NANTEN2 $^{12}$CO~($J$$=$2--1) data, we made integrated intensity maps for the blue and red-shifted  components in the spectrum as shown by the contours superposed on the VLA 20~cm radio continuum emission (Figure~\ref{fig: OutflowMap}).
The integrated velocity ranges of the blue and red-shifted components are $-15$ km~s$^{-1}$ to $+3$ km~s$^{-1}$ and $+15$ km~s$^{-1}$ to $+30$ km~s$^{-1}$, respectively.
The Mopra $^{12}$CO~($J$$=$1--0) data show the similar wing component for the blueshift side, but we could not find the red-shifted component possibly due to its slightly lower sensitivity than the NANTEN2 $^{12}$CO~($J$$=$2--1) data.
The obtained image clearly shows a presence of the strong CO emission from the Source A position, suggesting that the broad line emission is due to molecular outflow.  
With the same method described in Section~\ref{sec: SpectralTypeGasMass}, the molecular gas masses of the blue and red shifted components are derived to be 224 $M_{\odot}$ and 130 $M_{\odot}$, respectively.
The lower CO intensity and gas mass of the redshift component than the blueshift component can be attributed to the self absorption in the near side of the $+9$~km~s$^{-1}$ cloud. 
If we take into account the difference of the adopted distance to the clouds, this result is consistent with the gas masses obtained by \citet{Watson+07}, who gives 123 $M_{\odot}$ and 126 $M_{\odot}$ for the blue and red components, respectively.
In the same figure, we overlay another contours representing the integrated intensity of the NANTEN2 $^{12}$CO~($J$$=$1--0) data for the intermediate velocity cloud ($-1$~km~s$^{-1}$ $<$ $V_{\rm LSR}$ $<$ $+4$~km~s$^{-1}$), to compare the extension of the CO emission with the wing-like feature due to the outflow.
The Mopra $^{12}$CO~($J$$=$1--0) data also show a similar gas distribution at the 5$\sigma$ significance.
The low-density cloud at the intermediate velocity is more widely extended than the molecular outflows, suggesting that the broad line emission found in the spectrum and the position-velocity diagram includes the bridge component between the $-4$~km~s$^{-1}$ and $+9$~km~s$^{-1}$ clouds, in addition to the wing component originating from the massive stars.

\clearpage

\subsection{Triggering mechanisms of the high-mass star formation}
\label{sec: TrigMechanisum}

We found physical associations between the high-mass stars and the surrounding molecular gas separated into the three velocity components.
Here we discuss possible scenarios to form the high-mass stars, especially focusing on the triggering mechanisms introduced in Section~\ref{sec: Triggered high-mass star formation}.

\begin{figure}[h]
 \begin{center}
 \centering
  \includegraphics[width=15.0cm]{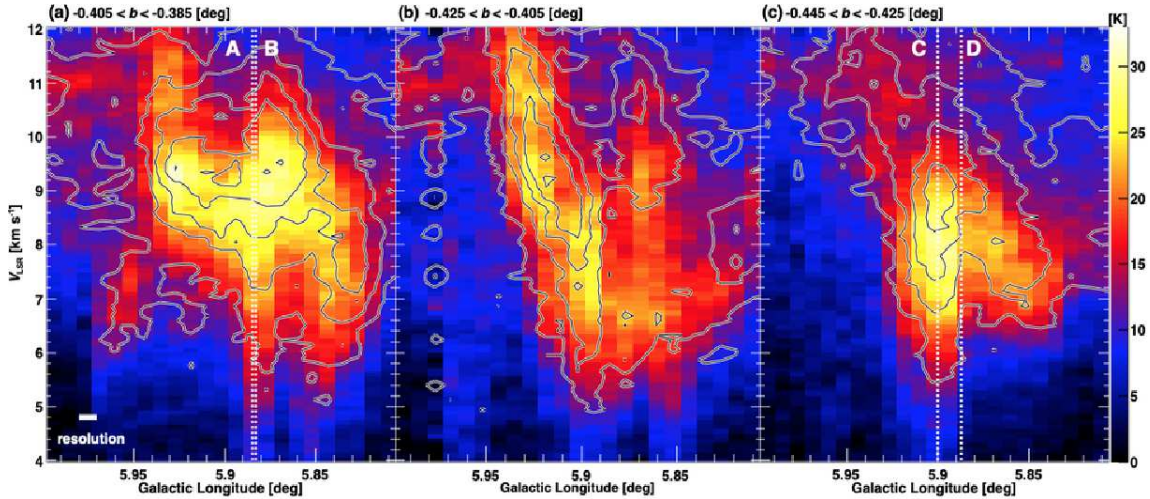}
  \end{center}
  \caption{Longitude-velocity diagram for the $+9$~km~s$^{-1}$ cloud obtained with the Mopra~$^{12}$CO~($J$$=$1--0) (image) and $^{13}$CO~($J$$=$1--0) (contour) data. The integrated latitude ranges are (a) \timeform{-0.D405} $< b <$ \timeform{-0.D385}, (b) \timeform{-0.D425} $< b <$ \timeform{-0.D405}, and (c) \timeform{-0.D445} $< b <$ \timeform{-0.D425} as shown in Figure~\ref{fig: Mopra13COChannelMap}. The contours are drawn by 3$\sigma$ step from 3$\sigma$ intensity level. The dotted line represent positions of the exciting sources A--D.} 
\label{fig: Mopra12CO13COLV}  
\end{figure}

\begin{figure}[h]
 \begin{center}
 \centering
  \includegraphics[width=16.0cm]{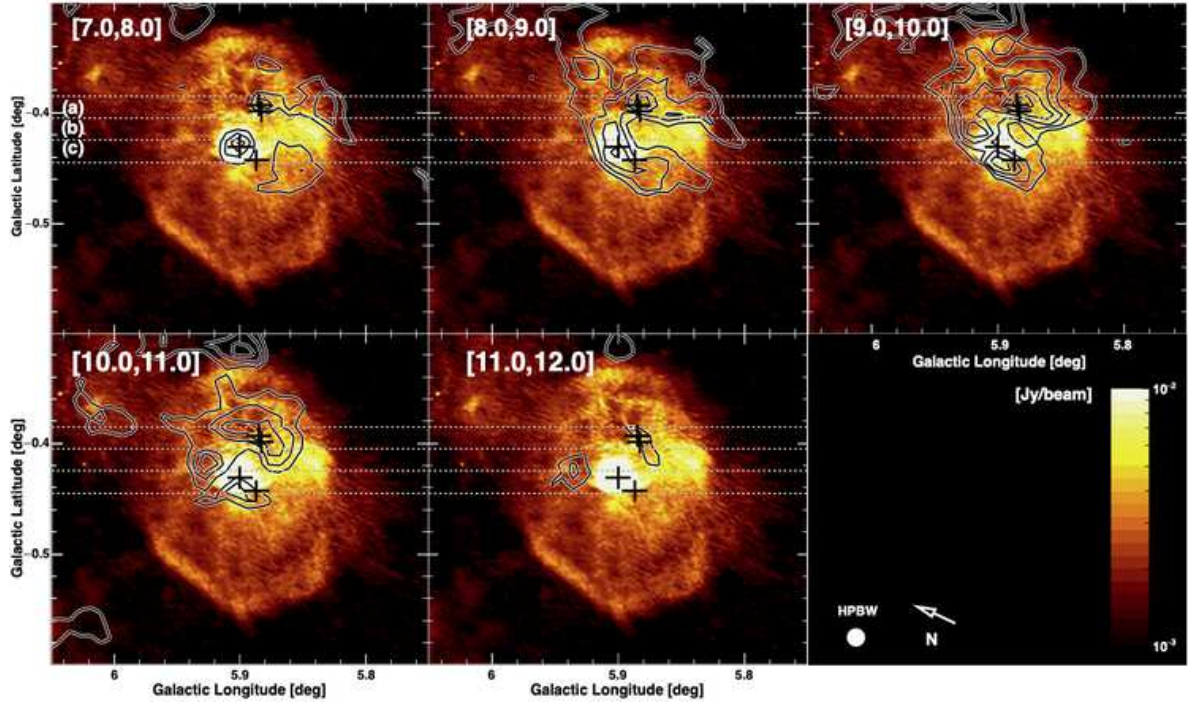}
  \end{center}
 \caption{Velocity channel map for the $+9$ km~s$^{-1}$ cloud using the Mopra $^{13}$CO~($J$$=$1--0) data (contour) compared to the VLA 20 cm radio continuum emission (image). The contours are drawn by 2$\sigma$ step from 3$\sigma$ level. The regions (a)-(c) separated by the horizontal dotted lines correspond to the integrated latitude range for the position-velocity diagrams shown in Figure~\ref{fig: Mopra12CO13COLV}. The crosses represent positions of the exciting sources A--D.}
\label{fig: Mopra13COChannelMap}  
\end{figure}

Figures~\ref{fig: Mopra12CO13COLV} (a)--(c) show longitude-velocity diagrams obtained by the Mopra~$^{12}$CO~($J$$=$1--0) data (image) and $^{13}$CO~($J$$=$1--0) data (contours) whose integrated latitude ranges are represented by the horizontal dotted lines in Figure~\ref{fig: Mopra13COChannelMap}, where the distributions of the Mopra~$^{13}$CO~($J$$=$1--0) (contours) and the VLA~20 cm radio continuum emission (image) are compared in a velocity channel map.
The panel (c) in Figure~\ref{fig: Mopra12CO13COLV}, showing the gas structure around the exciting sources C/D, mainly consists of the highest CO intensity peak with the wide velocity range toward the exciting source C, and clouds at $l =$ \timeform{5.D87} showing an elongated structure toward the south direction (see 8 km~s$^{-1}$ $< V_{\rm LSR}$ $<$ 9 km~s$^{-1}$ component in Figure~\ref{fig: Mopra13COChannelMap}). 
The panel (b) corresponds to the intermediate area between Sources C/D and A/B, having the velocity structure extended to the positive longitude direction and the relatively diffuse emission at $l =$ \timeform{5.D87}.
In the area of the diffuse gas component at $l =$ \timeform{5.D87}, strong radio continuum emission with the flow-like structure is detected, showing an anti-correlation with the gas component (see 7 km~s$^{-1}$ $< V_{\rm LSR}$ $<$ 10 km~s$^{-1}$ in Figure~\ref{fig: Mopra13COChannelMap}).
The panel (a), corresponding to the gas around sources A/B, has three peaks at $l =$ \timeform{5.D84}, \timeform{5.D88} and \timeform{5.D93}. 
Although physical association among these three components is not clear, their velocity structure (peak velocity and covering velocity range) is apparently different.
If the expanding gas motion by the UV radiation dominantly affects the gas distribution, these position-velocity diagrams expect to show circular structure as illustrated in Figure~8 in \citet{Torii+15}.
However, we do not find such circular structures in either diagram.
It is difficult to attribute the distinct multiple velocity components to the expanding motion of the \HII\ gas.
In addition, the expanding scenario can not explain the forming process of the first-born exciting star. 
We infer that the velocity difference among clouds are not produced by the expanding motions by the exciting stars but rather exists prior to the formation of the high-mass stars.

\begin{figure}[h]
 \begin{tabular}{cc}
  \begin{minipage}{0.5\hsize}
   \begin{center}
    \rotatebox{0}{\resizebox{8cm}{!}{\includegraphics{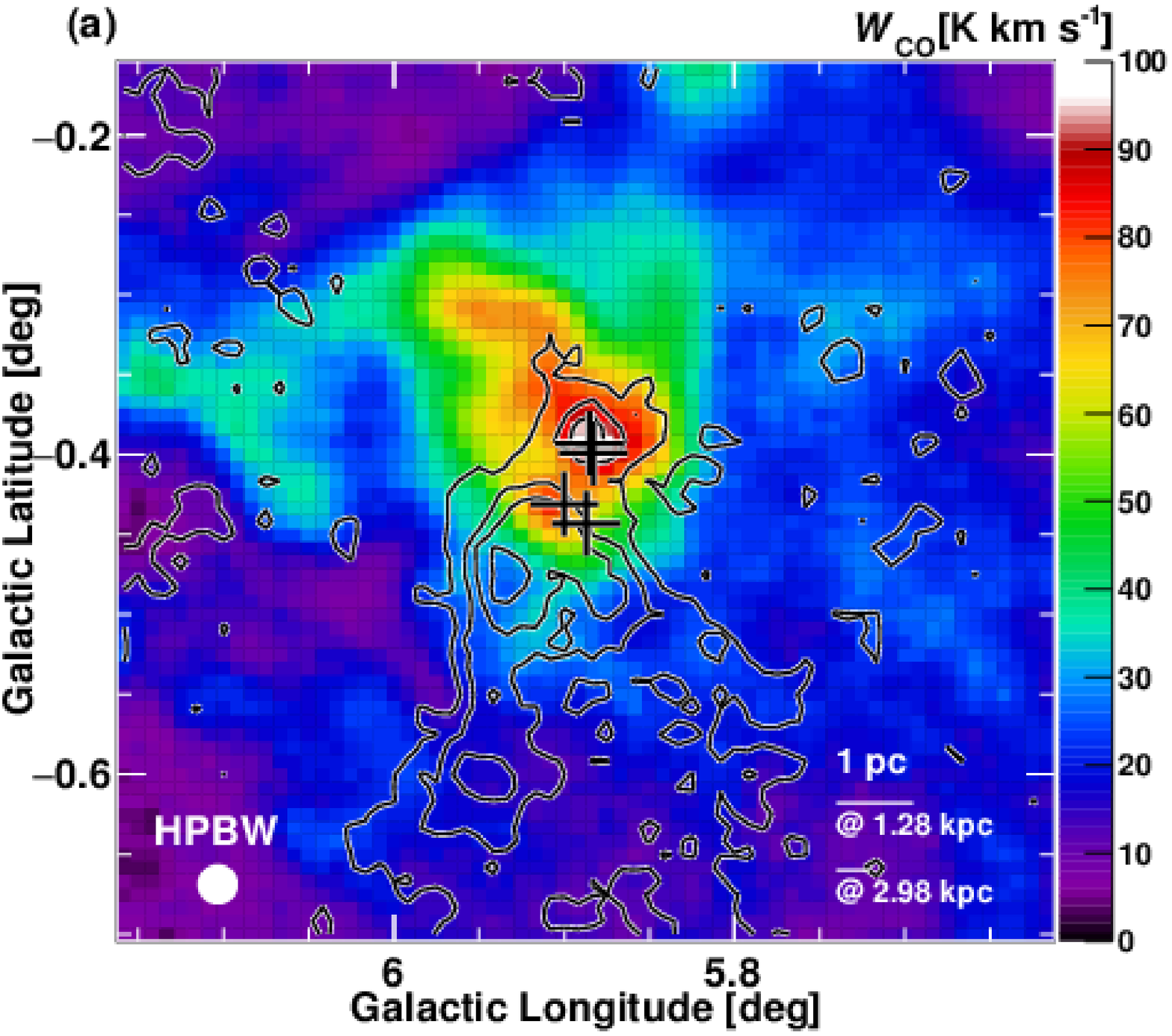}}}
   \end{center}
  \end{minipage} 
  \begin{minipage}{0.5\hsize}
   \begin{center}
    \rotatebox{0}{\resizebox{8cm}{!}{\includegraphics{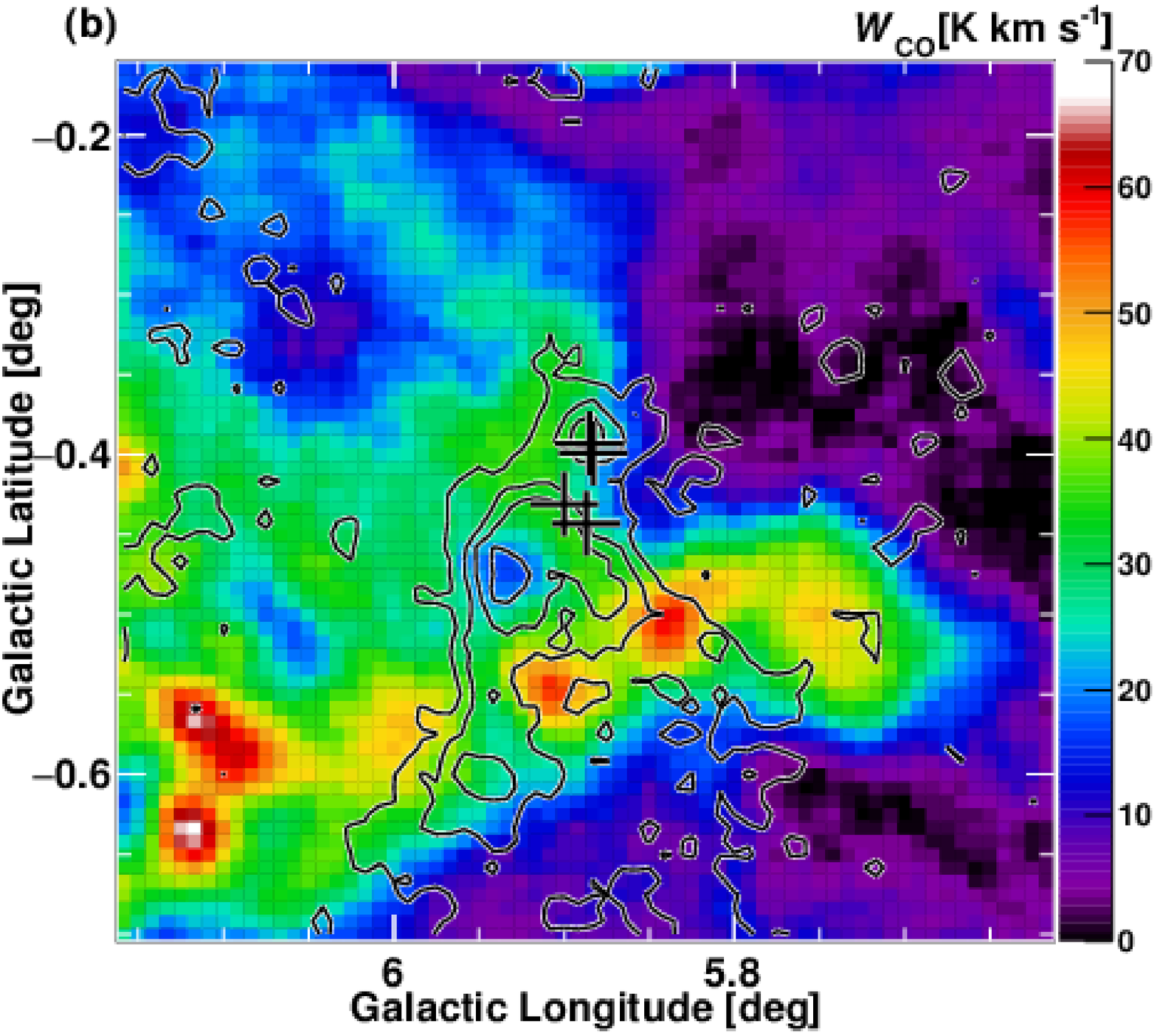}}}
   \end{center}
  \end{minipage} \\
  \end{tabular}  
  \caption{Comparisons of the two clouds using the velocity-integrated intensity maps of the NANTEN2 $^{12}$CO~($J$$=$2--1) data. (a) $-4$ km~s$^{-1}$ cloud (contour) and $+9$ km~s$^{-1}$ cloud (image). (b) $-4$ km~s$^{-1}$ cloud (contour) and $+16$ km~s$^{-1}$ cloud (image). The contour levels for the $-4$ km~s$^{-1}$ cloud are the same as Figure~\ref{fig: N212CO21compVLA20cmSpitzer8um}(a). The crosses represent positions of the exciting sources A--D.}
\label{fig: N212WCO21TwoClouds}   
\end{figure}

Figures~\ref{fig: N212WCO21TwoClouds}~(a) and (b) show comparisons of the velocity-integrated intensity maps of the NANTEN2 $^{12}$CO~($J$$=$2--1) data for the $-4$ km~s$^{-1}$ and $+9$ km~s$^{-1}$ clouds, and the $-4$ km~s$^{-1}$ and $+16$ km~s$^{-1}$ clouds, respectively. 
We found multiple complementary gas distributions such as toward $(l,~b) =$ (\timeform{5.D94}, \timeform{-0.D48}) and (\timeform{5.D95}, \timeform{-0.D60}); positions of the CO peaks in the $-4$ km~s$^{-1}$ cloud corresponds to the depression areas in the other clouds. 
The Mopra $^{12}$CO~($J$$=$1--0) data also showed similar gad distributions.
From table~\ref{table: gas masses}, the total gas mass of the $-4$~km~s$^{-1}$ and $+9$~km~s$^{-1}$ clouds in case of the distance of 1.28 kpc is $\sim$~3.8~$\times$10$^{4}$~$M_{\odot}$ and that of the $-4$~km~s$^{-1}$ and $+16$~km~s$^{-1}$ clouds is $\sim$~3.0~$\times$10$^{4}$~$M_{\odot}$.
The required gas mass ($M$) to bind the two clouds is estimated from the balance between the kinetic energy and the potential energy; $M = rv^{2}/2G$, where $v$ is the relative velocity between the two clouds, and $G$ is the gravitational constant.
Under an assumption of the cloud radius $r=5$~pc, these gas masses are calculated to be $M \sim$~9.8~$\times$10$^{4}$~$M_{\odot}$ and $M \sim$~2.3~$\times$10$^{5}$~$M_{\odot}$, respectively,
which are larger than the estimated total gas masses.
Even if we assume the distance is 2.98~kpc and the cloud radius is 11.5~pc (increasing proportionally to the distance), the tendency of the binding mass larger than the total gas mass does not change. 
These results indicate that the two clouds are not gravitational bound due to the large velocity difference up to $\sim$13~km~s$^{-1}$ and $\sim$20~km~s$^{-1}$.
Most of the gas in the $-4$~km~s$^{-1}$ and $+16$~km~s$^{-1}$ clouds is extended toward the negative latitude direction compared to the $+9$~km~s$^{-1}$ cloud, presumably outside the influence driven by sources A--D (see the contours in the position-velocity diagram of Figure~\ref{fig: N212CO21to10VB}), and thus it is difficult to expect that the simple expanding gas motion from the the $+9$~km~s$^{-1}$ cloud generates the observed gas structure.
These results suggest incidental collisions of the $-4$ km~s$^{-1}$ and $+9/+16$ km~s$^{-1}$ clouds rather than the expanding gas motion from the $+9$ km~s$^{-1}$ cloud.

The $-4$~km~s$^{-1}$ and $+9$~km~s$^{-1}$ clouds overlap mainly toward the regions with molecular gas around Sources A--D.
Whereas the $+9$~km~s$^{-1}$ cloud has presence of relatively high dense gas at $(l,~b)$ $\sim$ (\timeform{5.D93}, \timeform{-0.D30}), the $-4$~km~s$^{-1}$ cloud does not show significant CO emission from this area, where we do not find possible exciting sources.
These results are not inconsistent with the triggering scenario to form Sources A--D by the collision between the $-4$~km~s$^{-1}$ and $+9$ km~s$^{-1}$ clouds.
The bridging feature between the two clouds seen in Figure~\ref{fig: N212CO21LV} is a characteristic feature of the cloud-cloud collision as reported in many studies (see references in Section~\ref{sec: Triggered high-mass star formation}).
The high $^{12}$CO $J$$=$2--1 to 1--0 intensity ratio toward the exciting sources for the $-4$~km~s$^{-1}$ and $+9$ km~s$^{-1}$ clouds supports the physical correlation between the stars and the two clouds. 
Partly complementary gas distributions between the $-4$~km~s$^{-1}$ and $+9$~km~s$^{-1}$ clouds, and the $-4$~km~s$^{-1}$ and $+16$~km~s$^{-1}$ clouds, suggest physical interactions among these clouds.
On the other hand, we do not find a clear correlation between the $+9$~km~s$^{-1}$ and $+16$ km~s$^{-1}$ clouds from the cloud morphology. 
Although the $+16$ km~s$^{-1}$ cloud would be a part of the clouds forming the W28A2 region, it would be far away from the exciting stars compared to the other two clouds as discussed in Section~\ref{sec: PhysicalAssociationMCandStars}.

\begin{figure}[h]
 \begin{center}
 \centering
  \includegraphics[width=10.0cm]{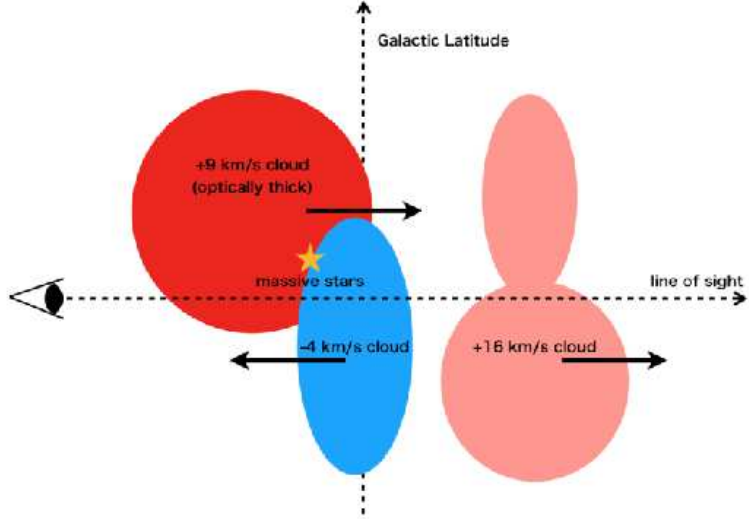}
  \end{center}
 \caption{Schematic image of the W28A2 region.}
\label{fig: W28A2_SchematicView}  
\end{figure}

From the above discussion, we propose that a cloud-cloud collision between the $-4$ km~s$^{-1}$ and $+9$ km~s$^{-1}$ clouds give a trigger the formation of the high-mass stars in the \HII\ region of W28A2.
Figure~\ref{fig: W28A2_SchematicView} shows a schematic image representing the configuration of the three velocity clouds and the massive stars in the W28A2 region.
A collision on the line of sight between the two clouds is occurring, giving a trigger for the formations of Sources A--D. 
The $+16$ km~s$^{-1}$ cloud is positioned at most far side from the exciting stars among the three clouds; the $+16$ km~s$^{-1}$ cloud partly shows a complementary gas distribution with the $-4$ km~s$^{-1}$ cloud as shown in Figure~\ref{fig: N212WCO21TwoClouds}(b) and thus expects a physical interaction in the past. 
The $-4$ km~s$^{-1}$ cloud is now colliding to the $+9$ km~s$^{-1}$ cloud just behind the high-density regions of the $+9$ km~s$^{-1}$ cloud.
This is consistent with a picture of the obscuration of the $+9$ km~s$^{-1}$ cloud, possibly due to the strong self-absorption toward the W28A2 region, as suggested from the spectral features obtained in this study and a previous molecular cloud observation \citep{Klaassen+06}. 

\subsection{Comparison with other candidates of the cloud-cloud collision}

Most of the candidates of the high-mass star or cluster formation trigged by the cloud-cloud collision show presence of complementary gas distribution, because the collision of a smaller cloud creates a hole in the larger cloud (\cite{HabeOhta92}; \cite{Takahira+14}). 
However, we do not find a clear complementary gas distribution toward the exciting sources A--D between the $-4$~km~s$^{-1}$ and $+9$~km~s$^{-1}$ clouds.
On the contrary, the CO intensities toward the sources A--D between the two clouds provides a positive intensity correlation (see Figure~\ref{fig: N212WCO21TwoClouds} (a)).
If the collision is an early stage (c.f., \cite{HabeOhta92}), a hole owing to the collision has not been created, and thus the complementary gas distribution may not be found.
The CO destruction by the UV radiation from the massive stars is not significant yet.
Such cloud properties have already been suggested for the Galactic super star cluster, RCW~38 \citep{Fukui+16} and NGC~6334 \citep{Fukui+18b}. 

RCW 38 is one of the youngest star forming region, where $\sim$20 O type stars are located. 
\citet{Fukui+16} suggested that the formation of high-mass stars in the cluster is triggered by the cloud collision about 0.1 Myr ago.
These massive stars are localized in the overlap region between the two clouds, which are connected by the bridging feature in the velocity space.
The rich gas around the massive stars suggests the ionization is not significant yet and the less complementary gas distribution implies the early stage of the collision.
This interpretation is also supported by \citet{Torii+19}, who suggested a young dynamical timescale ($<$ 1000 yr) for the star less cores in RCW~38 and the massive condensations in the cluster probably formed via cloud-cloud collision.
\citet{Fukui+18b} found the molecular gas toward the high-mass stars in NGC~6334 does not show a complementary gas distribution and has a positive intensity correlation between the colliding clouds.
The velocity difference between the colliding clouds is as high as $\sim$12~km~s$^{-1}$, which is similar to that of the $-4$~km~s$^{-1}$ and $+9$ km~s$^{-1}$ clouds in the W28A2 region.
Their large relative velocity indicates that the gas is accumulated by the collision in a short time, possibly leading to a rapid formation of the high-mass stars.

We also mention a possible correlation of the peak gas column density with the relative velocity between the colliding clouds and the number of massive stars for the regions triggered by the cloud-cloud collision \citep{Enokiya+19}. 
The peak column density for the W28A2 region obtained in this study is $\sim$4$\times$10$^{22}$~cm$^{-2}$ (for the $+9$~km~s$^{-1}$ cloud) and the relative velocity is $\sim$13~km~s$^{-1}$ (between the $-4$~km~s$^{-1}$ and $+9$ km~s$^{-1}$ clouds) if we assume the collision along the line of sight.
The W28A2 region holds at least four high-mass stars.
Our results are nearly consistent with their statistical study.
Even if we assume that the $+9$~km~s$^{-1}$ cloud has a few times larger column density obtained from the CS line observation \citep{Nicholas+12}, the consistency with the statistical study does not change.
These results support the triggered star-formation in the W28A2 region by the cloud-cloud collision.

\section{Summary}
\label{sec: summary}

We have investigated distributions and properties of the molecular gas in the \HII\ region W28A2 using the $^{12}$CO and $^{13}$CO~($J$$=$1--0) and $^{12}$CO~($J$$=$2--1) data obtained with the NANTEN2 and Mopra telescopes. 
The molecular clouds are extended within $\sim$5--10 pc from the center of \HII\ region, consisting of the three velocity components with the CO intensity peaks at $V_{\rm LSR} \sim$ $-4$~km~s$^{-1}$, 9~km~s$^{-1}$ and 16~km~s$^{-1}$.
The position-velocity diagram of the CO data show bridging features connecting the three clouds and coinciding with the directions of the exciting sources. 
Comparisons of the gas distributions with the radio continuum emission and 8 $\mu$m infrared emission show spatial coincidence/anti-coincidence, suggesting physical associations between the gas and the exciting sources.
The spectral type of the exciting stars are estimated to be O6--B0.5.

The obtained $^{12}$CO $J$$=$2--1 to 1--0 intensity ratio is $\gtrsim$~0.8 for the $-4$~km~s$^{-1}$ and $+9$~km~s$^{-1}$ clouds, suggesting physical associations with the the exciting sources.
At the CO intensity peak in the $+9$~km~s$^{-1}$ cloud, where the exciting stars are possibly embedded, the intensity ratio is relatively lower due to self absorption by the high-density clouds.
The position-velocity diagram does not show a feature expected in the case of the simple expanding gas motion from the exciting source.
The exciting sources are located toward the overlapping region of the $-4$~km~s$^{-1}$ and $+9$~km~s$^{-1}$ clouds and the bridging features between the two clouds are found toward this direction.
These gas properties are similar to the Galactic massive star clusters, RCW~38 and NGC~6334, where an early stage of cloud collision triggering the star formation is suggested.
The formation of high-mass stars in the \HII\ regions of W28A2 can be interpreted as a scenario, cloud-cloud collision.

\begin{ack}
NANTEN2 is an international collaboration of ten universities, Nagoya University, Osaka Prefecture University, University of Cologne, University of Bonn, Seoul National University, University of Chile, University of New South Wales, Macquarie University, University of Sydney, and Zurich Technical University.
The Mopra telescope is operated by the Australia Telescope National Facility. The University of New South Wales, the University of Adelaide, and the National Astronomical Observatory of Japan (NAOJ) Chile Observatory also supported the operations.

\end{ack}

\end{document}